\documentclass{aa}

\usepackage{graphicx}

\usepackage{txfonts}

\usepackage{amsmath}    
\usepackage{xcolor,colortbl}
\usepackage{soul}
\usepackage{xspace}
\usepackage{enumerate}
\usepackage{placeins}
\usepackage{makecell}
\usepackage[caption=false]{subfig}
\usepackage{pdflscape}
\usepackage{rotfloat}
\usepackage{adjustbox}
\usepackage{booktabs}
\usepackage{makecell}
\usepackage{multirow}
\usepackage{natbib}
\bibpunct{(}{)}{;}{a}{}{,}
\usepackage[pdftex,bookmarks=true,colorlinks=true,linkcolor=blue,citecolor=blue, urlcolor=blue,breaklinks]{hyperref}      
\usepackage{scalerel}
\usepackage{pgfplots}
\usepackage{tikz}
\usepackage{cuted}

\usetikzlibrary{svg.path}
\usetikzlibrary{shapes.geometric}

\definecolor{orcidlogocol}{HTML}{A6CE39}
\tikzset{
  orcidlogo/.pic={
    \fill[orcidlogocol] svg{M256,128c0,70.7-57.3,128-128,128C57.3,256,0,198.7,0,128C0,57.3,57.3,0,128,0C198.7,0,256,57.3,256,128z};
    \fill[white] svg{M86.3,186.2H70.9V79.1h15.4v48.4V186.2z}
                 svg{M108.9,79.1h41.6c39.6,0,57,28.3,57,53.6c0,27.5-21.5,53.6-56.8,53.6h-41.8V79.1z M124.3,172.4h24.5c34.9,0,42.9-26.5,42.9-39.7c0-21.5-13.7-39.7-43.7-39.7h-23.7V172.4z}
                 svg{M88.7,56.8c0,5.5-4.5,10.1-10.1,10.1c-5.6,0-10.1-4.6-10.1-10.1c0-5.6,4.5-10.1,10.1-10.1C84.2,46.7,88.7,51.3,88.7,56.8z};
  }
}

\newcommand\orcidicon[1]{\href{https://orcid.org/#1}{\mbox{\scalerel*{
\begin{tikzpicture}[yscale=-1,transform shape]
\pic{orcidlogo};
\end{tikzpicture}
}{|}}}}
\bibpunct{(}{)}{;}{a}{}{,}            

\pgfplotsset{compat=1.17}

\makeatletter
\renewcommand*\aa@pageof{, page \thepage{} of \pageref*{LastPage}}
\makeatother

\begin{document}

   \title{Feature-driven anomaly flagging in obscured active galactic nucleus light curves with autoencoders}
\titlerunning{Feature-driven anomaly flagging in obscured AGN light curves with autoencoders}

   \subtitle{}

   \author{Natale~De~Bonis \inst{1, 5, \orcidicon{0009-0001-1547-6648}}
          \and
          Demetra~De~Cicco \inst{2,1,3, \orcidicon{0000-0001-7208-5101}} \and
          Stefano~Cavuoti \inst{1, 4,\orcidicon{0000-0002-3787-4196}} \and
          Ylenia~Maruccia \inst{1, \orcidicon{0000-0003-1975-6310}} \and
          Dragana Ili\'{c}\inst{5, 6, \orcidicon{0000-0002-1134-4015}} \and
          Andjelka B.~Kova\v{c}evi\'{c}\inst{5, \orcidicon{0000-0001-5139-1978}}
          \and
          Giuseppe~Riccio \inst{1, \orcidicon{0000-0001-7020-1172}} \and
          Simone~Vaccaro \inst{1,\orcidicon{0009-0009-6526-4828}}
          }

   \institute{INAF - Astronomical Observatory of Capodimonte, Via Moiariello 16, I-80131 Napoli, Italy
         \and
             Department of Physics ``E. Pancini'', University Federico II of Napoli, Via Cinthia 21, I-80126 Napoli, Italy
        \and
            Millennium Institute of Astrophysics (MAS), Nuncio Monse\~nor Sotero Sanz 100, Providencia, Santiago, Chile
        \and
              INFN section of Naples, via Cinthia 6, I-80126, Napoli, Italy    
        \and 
            Department of Astronomy, Faculty of Mathematics, University of Belgrade, Studentski trg 16, 11000 Belgrade, Serbia
        \and 
            Hamburger Sternwarte,
            Universit\"{a}t Hamburg, Gojenbergsweg 112, D-21029 Hamburg, Germany}

   \date{Received Month XX, 2025; accepted Month XX, 2025}

  \abstract
 
   {Active galactic nuclei (AGN) are among the most complex classes of astrophysical objects, displaying a wide range of variability and observational properties. Identifying unusual AGN is crucial for understanding the physical mechanisms behind their emission better and for discovering potentially new subclasses or rare behaviors. With the increasing volume of data from next-generation surveys, machine-learning–based anomaly detection offers a promising approach to flagging and investigating such outliers systematically. }
   {We explore the use of unsupervised algorithms with a feature-driven approach to flag anomalous AGN, further explored by a human expert. The main focus is on obscured AGN, which tend to be harder to characterize.}
   {The algorithm we used was an AutoEncoder, which we trained on features extracted from the light curves rather than working with the light curves directly. The unsupervised nature of the method allows the detection of anomalies without relying on labeled data. To properly characterize the feature space and the detection process, we used the SHAP method.}
   {Our method flagged $11.18\%$ of the AGN we studied as anomalous.
   We focused in particular on anomalous obscured AGN and identified a refined subset of features that yields a comparable performance to the full set. Together with an in-depth analysis of the anomalies, this provides insight into how the AutoEncoder assigns anomalous status and which features are most indicative of astrophysically interesting behaviors or phenomena.}
   {}


   \maketitle

\section{Introduction}
We are now on the verge of great changes in the field of time-domain astronomy. 
With the advent of new-generation telescopes, which will cover the sky quickly and deeply, we will achieve a new look at the dynamic Universe. These changes will be led by the Legacy Survey of Space and Time (LSST; \citealt{lsstsciencecollaboration2009lsstsciencebookversion, 2019ApJ...873..111I}). The LSST will be conducted with the Simonyi Survey Telescope at the Vera C. Rubin Observatory. With LSST, our understanding of all the dynamical phenomena, such as variability in active galactic nuclei (AGN), will improve greatly, and hopefully, it will help us gain more insight into the physical process at play in the accretion process and place constraints on the structure of AGN. 
The primary survey of LSST, known as the Wide-Fast-Deep (WFD) survey, will cover 19.6k square degrees. We expect this region to be covered approximately 800 times over a decade, corresponding to an average cadence of about 3–4 days between visits. A portion of this time will also be dedicated to covering specific areas, known as deep drilling fields (DDF; \citealt{brandt2018activegalaxysciencelsst, scolnic2018optimizinglsstobservingstrategy}). In these regions, we already have information from previous surveys such as the Cosmic Evolution Survey (COSMOS; \citealt{Scoville_2007}). The observations of these fields aim to obtain a depth in the \textit{ugri} bands of $\sim28.5$ mag, $\sim28.0$ mag in the \textit{z} band, and $\sim 27.5$  mag in the  \textit{y} band, making them the ideal laboratories for AGN science. This is particularly relevant given the intrinsic variability of AGN at multiple timescales and wavelengths.

The volume of data produced by LSST will be unprecedented, with millions of sources observed each night. Therefore, it is crucial to develop methods that can help us analyze the enormous quantity of data produced quickly. To meet such challenges, we will use machine-learning (ML) algorithms. Many of these algorithms rely on supervised training, meaning that we need prior information about the data in the form of labels, that is, known outputs or classifications, so that the algorithm can learn to reproduce or predict them when presented with new, unseen data. The efficacy of the training process is strongly dependent on the features chosen. While other factors, such as the quality of the training data, the model architecture, and the optimization strategy, also play important roles, feature selection remains a critical step in ensuring that the model can generalize well and make accurate predictions. We need a feature set that captures the diversity of the population under study by sampling the parameter space broadly and uniformly. In practice, this is rarely achieved: observational biases, instrumental limitations, and selection effects often lead to incomplete or skewed coverage, which can compromise the model’s ability to generalize. Because of all these limitations, unsupervised models (i.e., algorithms that do not need previous knowledge) have gained more interest in recent years \citep{Fotopoulou_2024}. These techniques offer a great alternative in identifying patterns, groups, or anomalies in complex datasets. Specifically, the detection of anomalies (i.e., rare and unexpected objects that significantly deviate from the dominant population) offers a unique opportunity for scientific discovery, as they may indicate poorly understood or entirely new astrophysical phenomena \citep{2020MNRAS.498.3077W, 2022eas..conf..992M,2025AstTI...2...44O, iskandarli2026anomalyhunteralertsaha}. Moreover, anomaly detection can uncover systematic deviations arising from instrumental failures or calibration issues, thereby improving data quality and reliability \citep{Cavuoti_2024}. The automatic identification of such anomalies is essential, especially when dealing with massive datasets, because it provides a fast and scalable approach for filtering interesting cases, without the need for manual inspection.

We explore the use of unsupervised algorithms, specifically, an autoencoder (AE; \citeauthor{hinton1993autoencoders} \citeyear{hinton1993autoencoders}), for anomaly detection in AGN time series. An AE is a type of neural network designed to learn a compressed nonlinear representation of input data by training the model to reconstruct the original input from this lower-dimensional encoding. Their use is motivated by their ability to learn nonlinear representations of complex data distributions. Furthermore, AEs offer a flexible architecture, so we can easily tailor the neural networks to our specific needs. Last, their reconstruction-based anomaly classification helps us to explain the nature of the anomaly: by comparing the input time series with its reconstructed counterpart, we can identify which specific features the model fails to reproduce. {Although labels are available for a subset of the data, they are used exclusively for post hoc evaluation and interpretation of results, and never during training. The core detection framework therefore remains fully unsupervised.}

We use a feature-driven approach in which features extracted from the light curves serve as input, rather than working directly with the light curves, since most of the time AGN light curves vary in length due to differences in observational cadence, visibility windows, and data quality, making a direct comparison or modeling challenging. Moreover, feature-based representations allow us to incorporate domain knowledge explicitly, guiding the model toward aspects of the light curves that are known to be physically meaningful.

The paper is organized as follows. In Sect. \ref{dataset} we offer a brief description of the dataset. In Sect. \ref{method} we present our method. In Sect. \ref{results} we discuss the results, and finally, we draw our conclusions in Sect. \ref{conclusion}. 

\section{Dataset}
\label{dataset}
We used the same dataset as \cite{decicco19,De_Cicco_2021,refId0}, \cite{Cavuoti_2024}, and \cite{maruccia2025navigatingagnvariabilityselforganizing}. It consists of r-band observations from the three observing seasons of the COSMOS field by the VLT survey telescope (VST; \citeauthor{2011Msngr.146....2C} \citeyear{2011Msngr.146....2C}).
The three seasons cover a span of 3.3 years from December 2011 to March 2015 (more detailed information about the dataset can be found in  \cite{decicco19}).

The observations in the r band were originally planned to have a three-day cadence, depending on observational constraints. 
In practice, however, especially from the second observing season onward, the cadence was highly irregular, with two major gaps. The dataset spans three observing seasons, each separated by significant gaps in coverage. We acknowledge that the irregular cadence, combined with a minimum sampling of three days, limits the variability timescales that can be reliably probed with these light curves, it also limits the reliability of the damped random walk (DRW; \citealt{2009ApJ...698..895K, 2010ApJ...721.1014M}) parameters, and last, it can produce some artifact in the computed feature. Furthermore, this study is designed as a methodological foundation for future application to LSST deep drilling field data, where the improved cadence, depth, and ten-year baseline will allow these methods to reach their full potential. The depth, at each visit, was $\textit{r} \lesssim 24.6$ mag for point sources at a confidence level of $\sim 5\sigma$.

\begin{figure}
    \centering
    \includegraphics[width=1.0\linewidth]{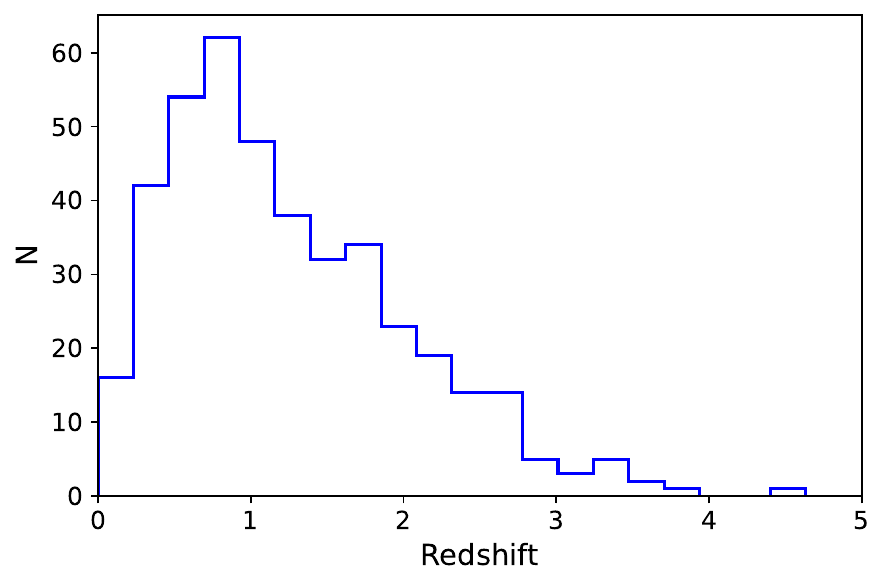}
    \caption{Redshift distribution of the AGN sample.}
    \label{fig:red}
\end{figure}

The dataset consists of 20,647 sources that were detected in at least $50\%$ of the visits, that is, 54, meaning that we only used light curves with at least 27 points, with an average magnitude in r $\leq$ 23.5 mag within a radius aperture of $1''$. For 2,414 of these sources, we know whether they are a star, a galaxy, or an AGN. In addition, for 413 of these sources, we also have a subclassification that determines the AGN type and the selection criterion, that is, whether they were detected in X-ray or the medium-infrared (MIR), or were selected based on variability. This was already used in \cite{De_Cicco_2021} and \cite{refId0}. Fig. \ref{fig:red} shows their redshift distribution. The distribution peaks at $z\sim1$ and extends to $z\sim4.5$, with the majority of sources lying at $z<2$. The sources were thus sorted into the various categories: 225 unobscured AGN, 122 obscured, 359 X-ray selected, 225 MIR selected, and 259 variability selected. The details of the selection criteria can be found in Sect. 2.2 of \citet{De_Cicco_2025}. Some sources are classified by multiple diagnostics. 
The whole set of features we used to characterize the sources is listed in Table \ref{tab:variability_features}. Some features were derived through the Feature Analysis for Time Series (FATS) Python library \citep{nun2015fatsfeatureanalysistime}, some others are DRW fit parameters, and last, we also have one morphological feature. Fig. \ref{fig:fetaure_comparison} shows the feature distribution for the labeled set (LS) and the unsupervised sample (US).
We concentrated on studying the AGN sample defined above. The main focus was on the obscured AGN, as their optical emission is more difficult to detect. As is widely accepted today, the central engine of an AGN is characterized by a disk-like structure, where matter is accreted onto a supermassive black hole, producing intense UV and optical emission. According to the unified model of AGN \citep{Antonucci1993-bb, Urry_1995}, the observed properties of an AGN primarily depend on its orientation with respect to our line of sight. In this framework, unobscured AGN are viewed face-on, allowing us a direct view of the accretion disk and broad-line region, while obscured AGN are observed edge-on. In the latter case, the central regions are obscured by a dusty toroidal structure (the torus) that efficiently absorbs high-energy photons and obscures the UV/optical emission from the accretion disk. The observational limitations inherent to obscured AGN, most notably, the obscuration of the central engine by the dusty torus, make it challenging to characterize their intrinsic variability using standard photometric diagnostics. In this context, unsupervised techniques offer a powerful approach, as they do not rely on explicit labels and can identify subtle deviations from the expected behavior learned from the majority population.

\section{Methods}
\label{method}
In this section, we describe the method we adopted to detect anomalies in AGN light curves.
The core of our approach relies on an AE \citep{hinton1993autoencoders}, an unsupervised neural network trained to reconstruct its input through a compressed latent representation. 
We first outline the architecture and training procedure of the AE, and we then describe the anomaly detection strategy and the methods we used to interpret the model outputs.

\subsection{Autoencoder architecture}
\begin{figure*}
    \centering
    \includegraphics[width=1.0\linewidth, trim=0 90 0 0, clip]{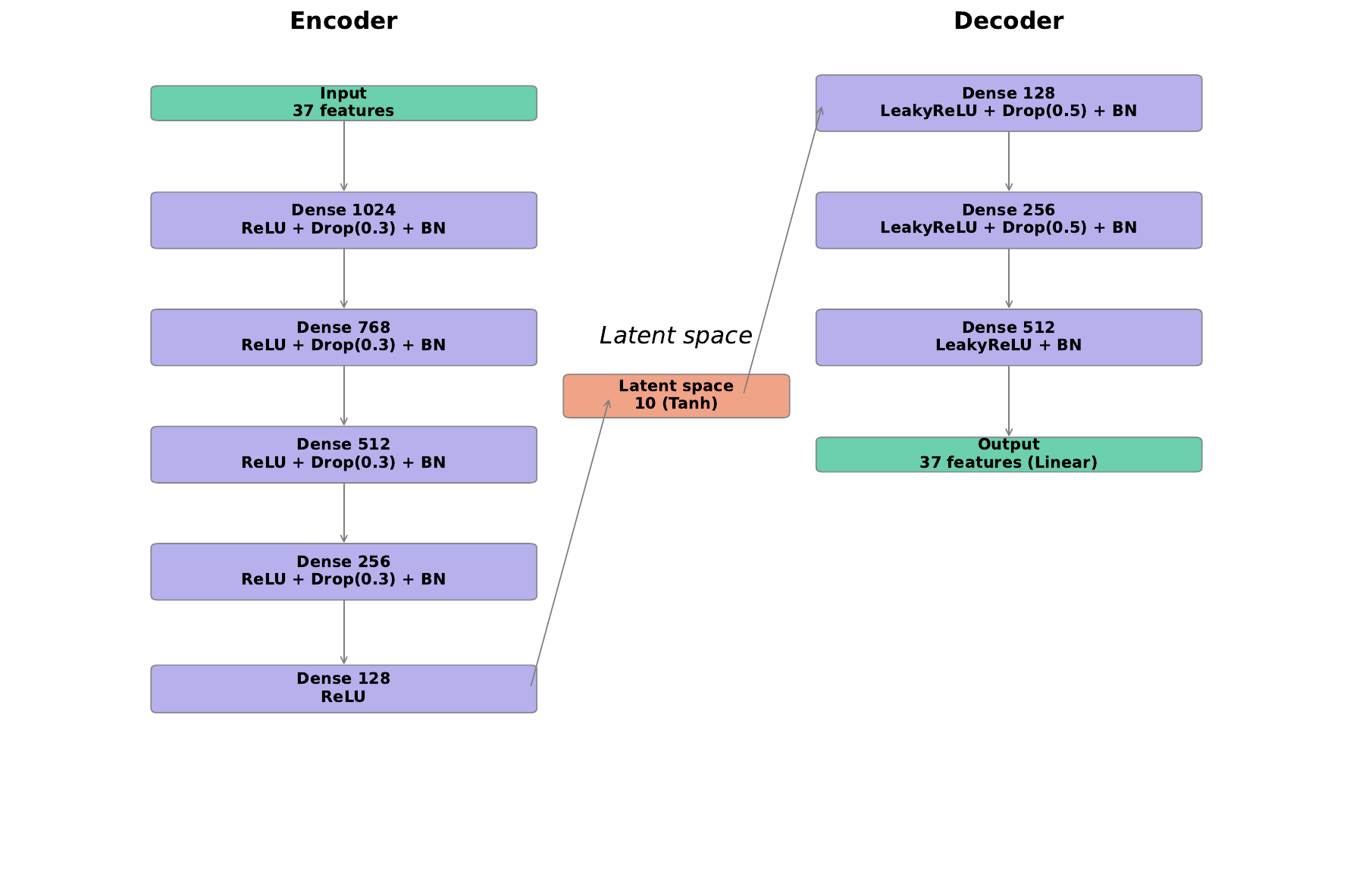}
    \caption{\centering Visual representation of the AutoEncoder architecture.}
    \label{fig:AE_arch}
\end{figure*}

Autoencoders are neural networks that learn a nonlinear mapping from input to output through an intermediate low-dimensional latent space. 
They consist of two parts: an encoder, which compresses the input data, and a decoder, which reconstructs the input from this compressed representation. 
The AE we used is composed of fully connected layers with nonlinear activations and regularization mechanisms, as summarized in Fig.~\ref{fig:AE_arch}.
To reduce the risk of overfitting\footnote{Overfitting is a common issue. The model learns to reproduce the training data too precisely and fails to generalize to new inputs.}, we incorporated dropout layers, which randomly deactivate a fraction of the neurons during training. We used a dropout rate of 0.3, meaning that $30\%$ of the neurons were ignored at each training step. This forces the network to learn more robust and distributed representations. We also applied batch normalization layers, which standardize the inputs to each layer during training. This helps to stabilize and accelerate the learning process by reducing the internal covariate shift.
The encoder progressively reduced the input dimensionality starting from 37 features by mapping it into a ten-dimensional latent space. Each dense layer applied a nonlinear activation called rectified linear unit ( ReLU) to capture complex patterns in the data. The ReLU is defined as follows:
\begin{equation}
\text{ReLU}(x) = \max(0, x).
\end{equation}
The decoder reconstructed the original input from the latent representation, using similar techniques but in reverse, with leaky rectified linear unit (LeakyReLU) activations that are defined as follows:
\begin{equation}
\text{LeakyReLU}(x) = \max(\alpha x, x),
\end{equation}
where $\alpha$ is a small positive constant, in our case, $\alpha=0.1$.
Last, a linear output layer was used to match the input scale. To ensure consistent input in all samples, we preliminarily scaled each feature using the StandardScaler implementation from the scikit-learn library. This method scales each feature so that it has zero mean and unit variance. This makes the training process more efficient and prevents features with wider numerical ranges from dominating the learning dynamics.
As a loss function, we minimized the mean squared error (MSE) using the Adam \citep{kingma2017adammethodstochasticoptimization} optimizer with a learning rate of 1e-4 and a batch size of 32. Gradient clipping with a maximum norm of 1.0 was applied to stabilize training.
\subsection{Anomaly-detection strategy}
The logic behind the use of the autoencoder for anomaly detection lies in its ability to reconstruct the input data accurately when the data resemble those seen during training.  As a consequence, if a light curve shows a large reconstruction error (RE), it is flagged as anomalous. The reconstruction error was computed as the mean squared error (MSE), defined as follows:
\begin{equation}
    \textbf{RE}= \frac{1}{d} \sum_{j=1}^{d} (x_{ij} - \hat{x}_{ij})^2,
\end{equation}
where \( d \) denotes the dimensionality of the feature space, and $x_{ij}$ and $\hat{x}_{ij}$ represent the reconstructed and target feature value for the ith time series. 

The anomaly detection threshold was defined during the training as the $95^{\rm th}$ percentile of the RE distribution of the US. So, a light curve was flagged as anomalous when its RE was larger than the threshold defined above. 
To examine the spatial distribution of this point compared to the normal ones, we used a T-distributed stochastic neighbor embedding (t-SNE; \citealt{maaten2008visualizing}).  t-SNE is a nonlinear dimensionality reduction
technique that maps high-dimensional data into a two- or three-dimensional space, preserving local neighborhood structures. Points that were similar in the original space are mapped closely together, whereas dissimilar points are mapped farther apart. 

We stress that we used the t-SNE purely as a qualitative visualization tool, and no quantitative conclusions were drawn from the resulting projections.
\subsection{Quantifying feature impact on the anomaly detection}
To assess the AE capability to flag anomalous AGN, we need to understand which features contributed most to the observed RE. However, since we worked with an unsupervised algorithm, it was not straightforward to interpret the results. To overcome this limitation, we trained a light-gradient boosting machine regressor (LGBMRegressor; \citealt{LightGBM_Paper2017}) as a surrogate model, which can approximate the mapping between input features and RE values. LightGBM is a gradient-boosting algorithm based on decision trees that adopts a leaf-wise growth strategy, granting high scalability, reduced computational cost, and strong predictive performance in regression tasks.
To interpret the output of the surrogate model, we employed the Shapley additive explanations (SHAP) framework \citep{NIPS2017_7062}. SHAP assigns each feature a contribution value, known as a SHAP value, based on its marginal impact on the model prediction. These values are computed by considering all possible combinations of feature subsets and measuring how the inclusion of a given feature changes the output. We used the TreeExplainer module from the SHAP library, which is specifically optimized for tree-based models such as LightGBM. For each data point, SHAP provides a vector of feature attributions indicating how much each feature increased or decreased the predicted RE, with higher absolute SHAP values indicating greater effect. We first applied this framework to the full LS to assess which features drive the reconstruction error in all source classes, and we then focused specifically on the AGN population.

\begin{figure}
    \centering
    \includegraphics[width=1.0\linewidth]{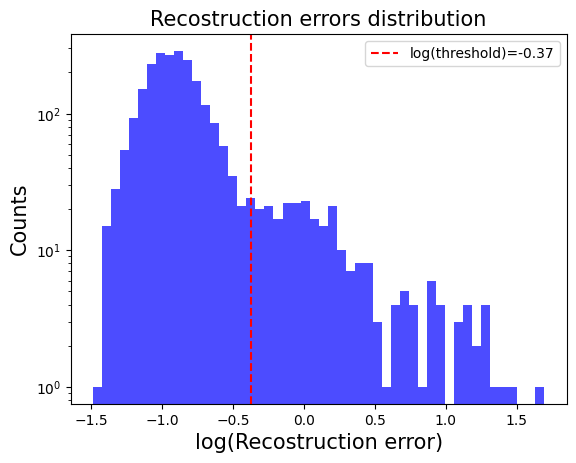}
    \includegraphics[width=1.0\linewidth]{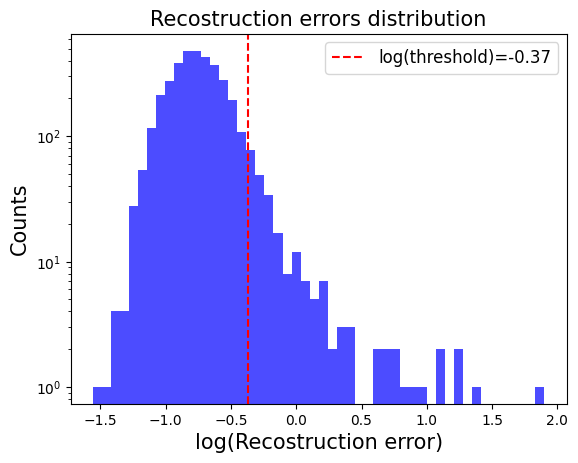}
    \caption{Comparison between the reconstruction error distributions of the LS (\textit{top}) and the US (\textit{bottom}). The dashed red line in both panels marks the anomaly detection threshold defined during training ($95^{\rm th}$ percentile of the US distribution).}
    \label{fig:rec_tot}
\end{figure}

\begin{figure}
    \centering
    \includegraphics[width=1.0\linewidth]{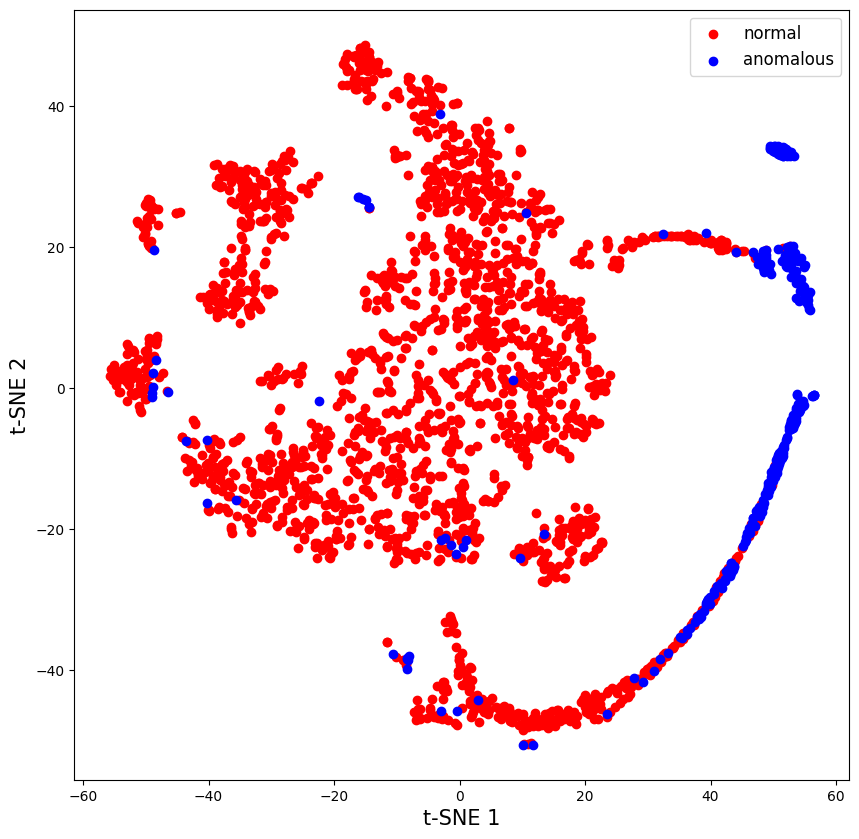}
    \caption{t-SNE analysis of LS in which we highlight the normal and anomalous data. The red points represent the normal data, and the blue points represent the anomalous data. The two classes are mostly well separated.}
    \label{fig:tsne_tot}
\end{figure}

\begin{figure*}
    \centering
    \includegraphics[width=1.0\linewidth]{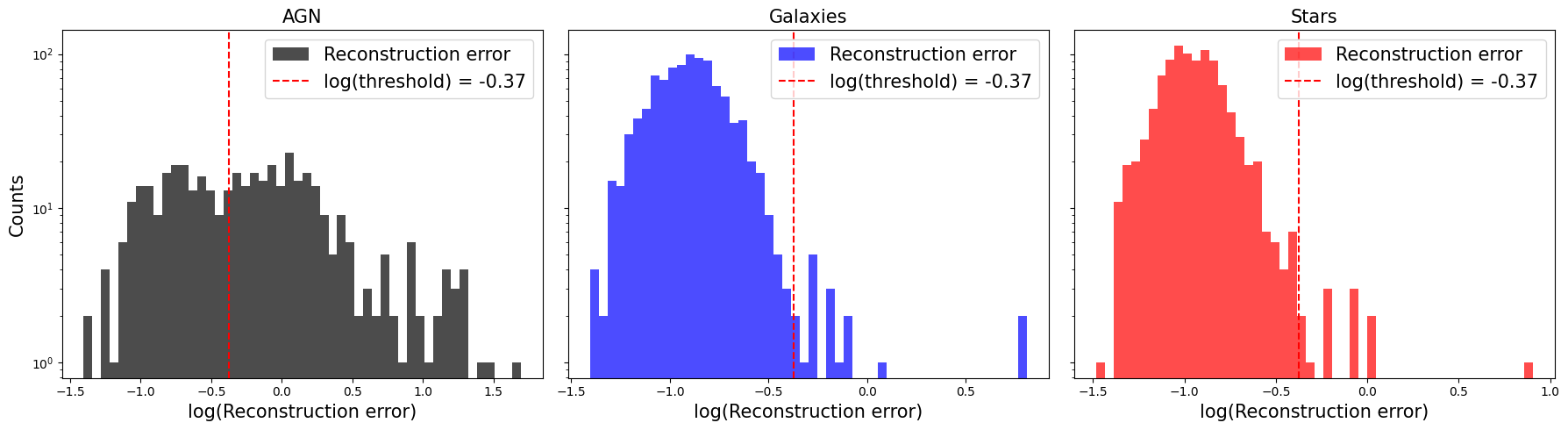}
    \caption{Reconstruction error of the label set subdivided into classes: AGN (\textit{left}), galaxies (\textit{center}), and stars (\textit{right}).}
    \label{fig:rec_classes}
\end{figure*}

\begin{figure}
    \centering
    \includegraphics[width=1.0\linewidth]{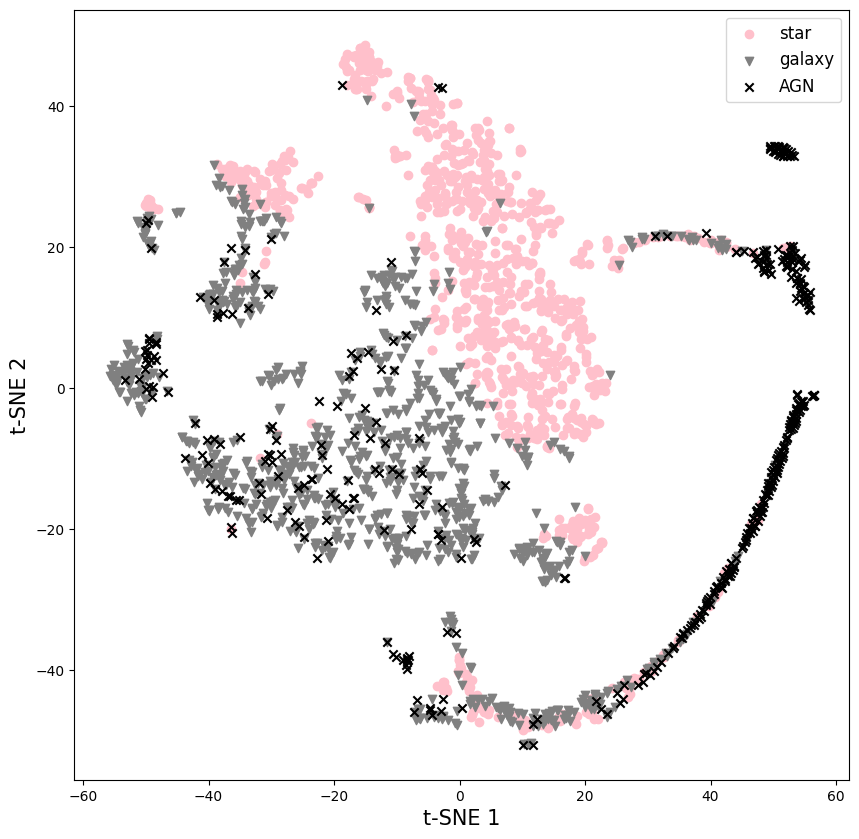}
    \caption{t-SNE analysis of the LS in which we highlight the various classes. The black crosses represent the AGN, the pink dots represent the stars, and the gray triangles represent the galaxies.}
    \label{fig:classes_tsne}
\end{figure}

\begin{figure}
    \centering
    \includegraphics[width=1.0\linewidth]{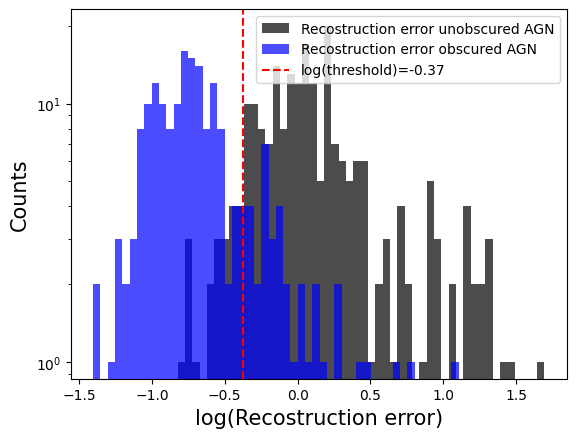}
    \caption{Reconstruction error distribution for the AGN population, in which we highlight the classification in unobscured and obscured AGN.}
    \label{fig:re_1_2}
\end{figure}
To quantify the improvement, if there was any, of using the most discriminative feature, we trained two models: a random forest classifier \citep{2001MachL..45....5B}, and a GradientBoostingClassifier \citep{friedman2001greedy}, both from the Scikit-learn Python library. The random forest algorithm builds an ensemble of decision trees on randomly selected subsets of the data and features and aggregates their predictions through majority voting. In contrast, gradient boosting constructs trees sequentially, where each new tree is trained to correct the errors made by the previous ones, optimizing a loss function via gradient descent. We trained these algorithms on both feature sets and evaluated the performance using a stratified five-fold cross-validation. In a stratified five-fold cross-validation, the dataset is split into five subsets, each maintaining the original proportion of normal and anomalous samples. The model is trained five times, each time using four folds for training and one for validation. This ensures more robust and unbiased performance estimates, in particular, for small or imbalanced datasets, such as in our case. At each of the five iterations, we computed the following three metrics: 
\begin{itemize}
    \item {Receiver operating characteristic area under the curve (ROC AUC):} this quantifies the model ability to distinguish between classes at varying decision thresholds.
    \item{Precision:} the proportion of true positive predictions among all predicted positives.
    \item{Recall:} the proportion of true positive predictions among all actual positive instances.
\end{itemize}

\section{Results}
\label{results}
The distribution of RE of the LS is shown in the top panel of Fig. \ref{fig:rec_tot}. The percentage of light curves beyond the threshold is $11.18\%$. This is more than double of what we observed in the US (shown in the bottom panel of the same figure). This is expected since the composition of the two samples is different, with the US being more enriched in AGN, which are known to exhibit a more complex and variable light curve behavior.
Fig. \ref{fig:tsne_tot} shows that from the t-SNE representation of the data, the anomalous points tend to be mostly found in the outer region of the distribution. 
We then used the classes of the LS to determine how the AE reconstructed each class (star, galaxies, and AGN). Figure \ref{fig:rec_classes} shows the distribution of the reconstruction error for each class. {A full overview of the flagged sources is listed in Table  \ref{tab:anomaly_summary}. Most of the objects flagged as anomalies are AGN. To be precise, 242 out of 272 detected anomalies are AGN. At the same time, only 16 galaxies and 14 stars are flagged as anomalous. This highlights that it is hard for the AE to reconstruct the stochastic behavior of AGN, which significantly deviates from the patterns learned during training on the bulk population, and which reflects the different composition of the LS compared to the US. Within the AGN population, the flagging rate varied strongly with classification: unobscured AGN were flagged at a rate of $90.7\%$, while obscured and unclassified AGN were flagged at $31.1\%$ and $1.5\%$, respectively. This is also shown in the t-SNE analysis in Fig. \ref{fig:classes_tsne}, in which we highlight the various classes, and which also shows, as expected, that the classes are not perfectly separated.
\begin{table}[h]
\centering
\caption{Anomaly detection results.}
\begin{tabular}{lccc}
\hline\hline
Class & Total & Flagged & Fraction (\%) \\
\hline
Unobscured AGN & 225 & 204 & 90.7 \\
Obscured AGN   & 122 & 38  & 31.1 \\
Unclassified AGN & 66 & 1  & 1.5  \\
Galaxies       & 1000 & 16 & 1.6  \\
Stars          & 1000 & 14 & 1.4  \\
\hline
\end{tabular}
\tablefoot{The flagged fraction represents the proportion of sources in each class identified as anomalous by the AE.}
\label{tab:anomaly_summary}
\end{table}
For the AGN sample in particular, when we split the RE of the AGN class, according to the subclassification into unobscured and obscured AGN, for sources with an available subclassification, we can obtain a major insight into how the AE handled these sources. Fig. \ref{fig:re_1_2} shows that the majority of the points that were flagged as anomalous are unobscured AGN. This is an expected result because of their unobscured accretion disks and inherently stochastic variability. Specifically, 204 of the 242 anomalous AGN were classified as unobscured, while only 38 correspond to obscured AGN. This marked imbalance suggests that the model struggles more with reconstructing the variability in unobscured AGN. Unobscured AGN, which offer a direct view of the accretion disk and broad-line region, exhibit a rich and stochastic variability that deviates significantly from the patterns learned during training. In contrast, obscured AGN are obscured by a dusty torus, and their observed light curves are dominated by host galaxy emission. As a result, their variability resembles that of normal galaxies.
These results are further supported by the SHAP analysis performed on the full LS. As shown in Fig.~\ref{fig:shap_class}, the features driving the anomaly detection differ across source classes. For anomalous AGN, the flagging is primarily driven by variability features such as \texttt{ExcessVar}, \texttt{GP\_DRW\_sigma}, and \texttt{Meanvariance}, which are directly linked to the stochastic variability of the accretion disk. In contrast, anomalous galaxies and stars are dominated by the structure function slope \texttt{SF\_ML\_gamma}, indicating a different origin for their anomalous classification. This suggests that the AE is sensitive to AGN-specific variability properties. These results are complemented by Fig.~\ref{fig:shap_agn}, which compares the mean absolute SHAP values for anomalous and normal AGN separately. Only 7 out of 15 features are shared between the two groups, and notably, the features dominating the anomalous AGN list, \texttt{ExcessVar}, \texttt{GP\_DRW\_sigma}, and \texttt{Meanvariance}, are absent from the normal AGN top features. Even the effect of the common features on the RE is weaker for normal sources. All these findings indicate that the flagged AGN are genuine outliers within their class, driven by physically motivated variability properties.

Not all of these flagged AGN or any other labeled source flagged as anomalous are necessarily physically meaningful; some of these detections are surely due to systematic artifacts (e.g., a gap that is too large, or poor sampling). For this, the model aim is not to give a definite answer as to the anomaly status of a source, but to highlight sources that show atypical behaviors as compared to the overall population. For this, the use of SHAP is extremely important as it helps the user to understand why these sources stand out to the AE, and to help the user in the final inspection and selection of the sources that are worth following. A more detailed discussion of the physical versus systematic nature of the detected anomalies is provided in Sect.~\ref{sec:obscured}, where we focus specifically on the obscured-AGN population, for which the SHAP analysis enables a source-by-source characterization of the drivers behind each anomalous detection.
\begin{figure*}
    \centering
    \includegraphics[width=1.0\linewidth]{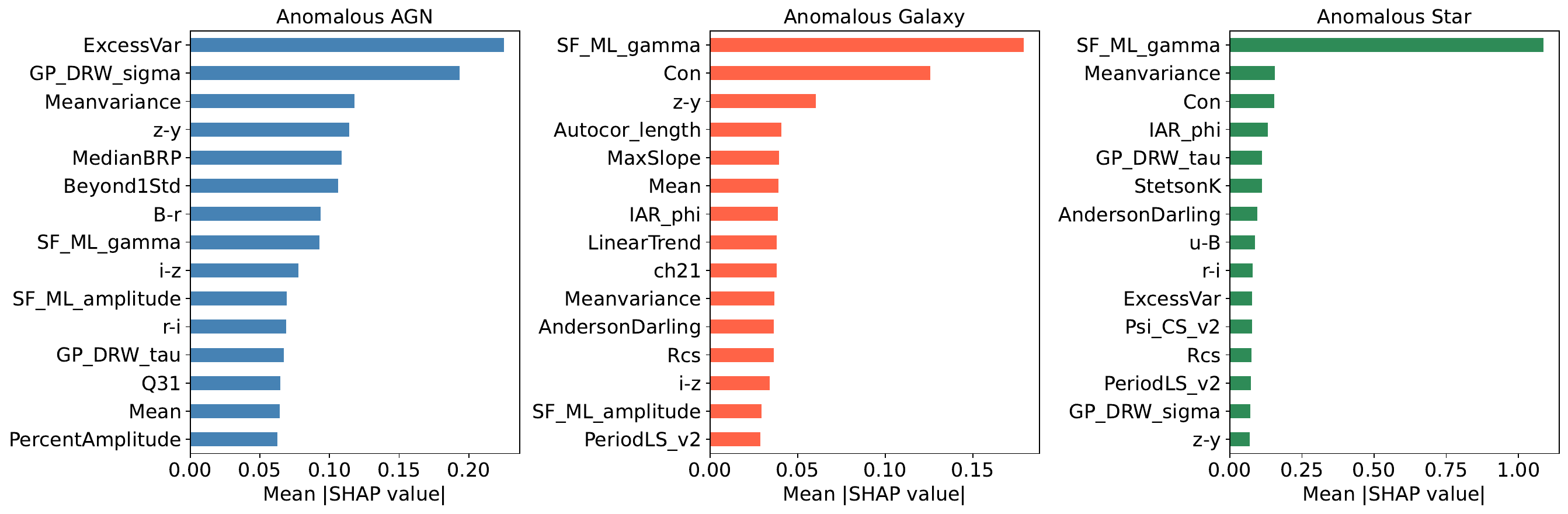}
    \caption{Mean absolute SHAP values for the top 15 most influential features, computed for anomalous sources in each labeled class: AGN (\textit{left}), galaxies (\textit{center}), and stars (\textit{right}).}
    \label{fig:shap_class}
\end{figure*}
\begin{figure}
    \centering
    \includegraphics[width=1.0\linewidth]{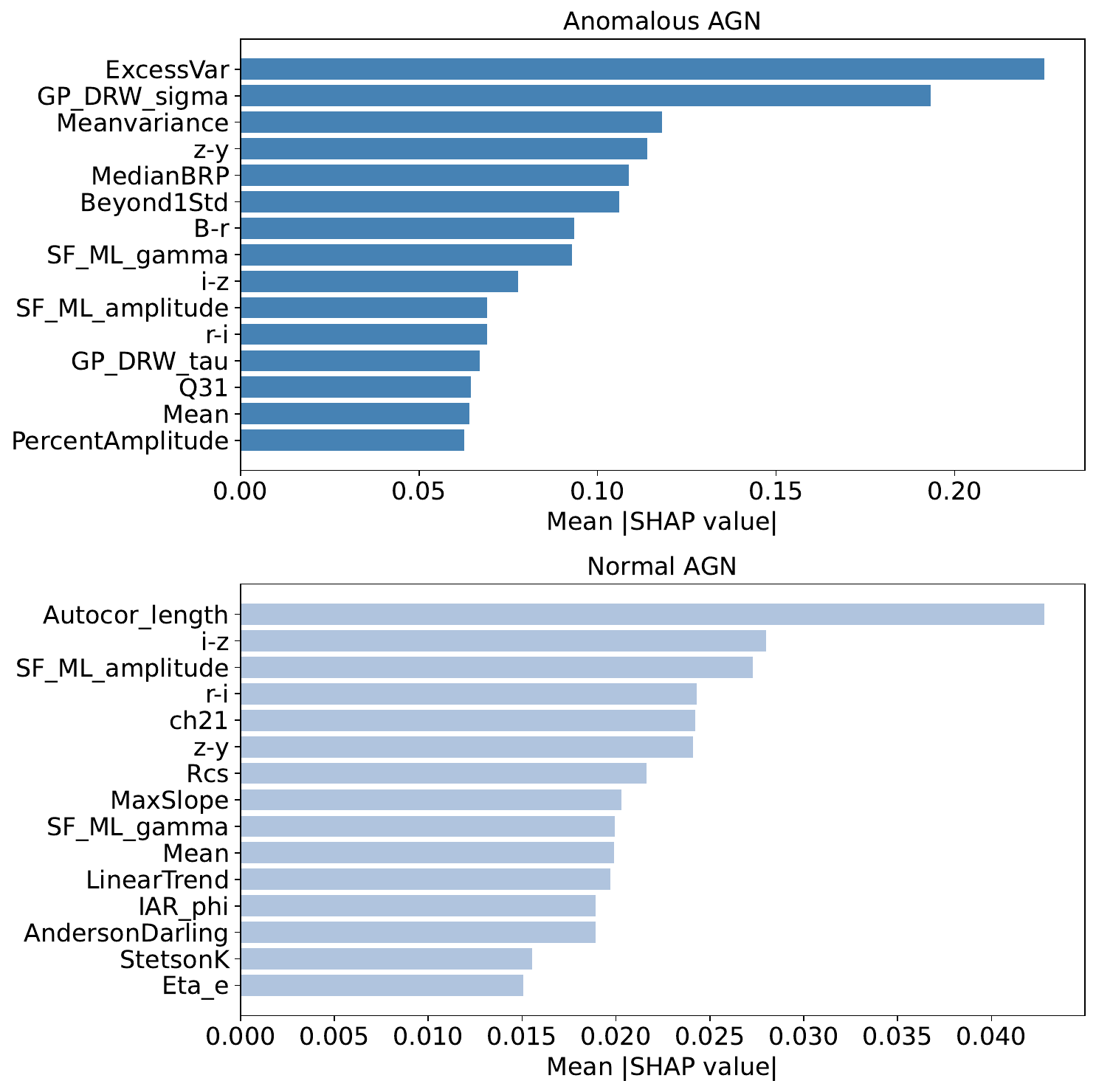}
    \caption{Comparison between the mean absolute SHAP values for the top 15 most influential features for flagged and non-flagged AGN. The two panels have different x-axis scales to better emphasize the scale of the magnitude difference between the feature effects.}
    \label{fig:shap_agn}
\end{figure}
\subsection{Obscured AGN}
As previously discussed, obscured AGN are notoriously difficult to distinguish from normal galaxies based on their optical variability properties alone. For this reason, it is particularly valuable to investigate whether the features contributing most to a given anomalous reconstruction are located in the tails or in the central regions of their respective distributions. If a highly influential feature for an anomalous obscured AGN falls within the end of its distribution (i.e., in the tails), this could give us information about strongly atypical behavior. Moreover, the identification of specific features that are extreme can provide more concrete explanations for the flagging of a particular source as anomalous. On the other hand, if the most contributing features lie closer to the center of the distribution, the anomaly may be the result of a subtler interplay of features and not a single extreme value. So, this type of analysis can help us to uncover the working of the anomaly detection mechanism and to characterize the detected anomalies.
However, as mentioned above, our primary interest lay in anomalous obscured AGN because they are significantly more challenging to distinguish from normal galaxies based on their optical variability characteristics alone. The AE classification of these sources as anomalous may result from several factors, including contamination by intermediate types, insufficient sampling, or photometric inconsistencies. A detailed analysis of the behavior of these obscured AGN can therefore help us understand how the AE identifies these anomalies, and more broadly, it might provide insight into the intrinsic diversity of the obscured AGN population.

In the following, we label points as tail points whose most influential feature lies in the tail of its distribution (i.e., anomalous sources whose most influential feature lies in the tails of its distribution: $<5^{\rm th}$ percentile or RE $>95^{\rm th}$ percentile), and non-tail points are those whose most influential feature is not in the tail of its distribution. Of the 38 detected anomalous obscured AGN, 14 are tail points and 24 are non-tail points.

We further examined the ranked feature importance of the 24 non-tail points to determine the first feature in the list that falls in the tails, if there was one. We refer to this as the leading tail feature, defined as the highest-ranked feature (according to SHAP importance) whose value falls outside the central $90\%$ interval of its distribution in the full dataset. We ordered them by their number of appearances, and they are shown in Fig. \ref{fig:feature_tot}. Although these features were not the top contributors in all cases, their consistent appearance as the most deviant from the standard suggests that they play a key role in defining the boundary between typical and atypical behavior for obscured AGN. This subset of features, composed of 19 out of the 37 starting features, can be thought of as a set of features that is very good at separating anomalous and normal obscured AGN.  To ensure that each feature was able to distinguish between normal and anomalous points, we compared their probability density functions (PDFs) using the Kolmogorov-Smirnov (KS) test \citep{1958ArM.....3..469H}. Only features with a statistically significant divergence (e.g., a p value $< 0.05$) were retained for further analysis.

Figure \ref{fig:feature_pdf} illustrates the PDFs of the original 19 candidate features. Finally, to validate the effectiveness of this reduced feature set, we proceeded with visual and quantitative assessments of the class separability.
First, we employed t-SNE to project the data into a two-dimensional space and compared the separation between normal and anomalous samples for the whole feature set and for the 12 most discriminative features. The results of this visual comparison are shown in Fig. \ref{fig:comparison_tsne}. The t-SNE plots indicate a clearer boundary between the two groups for the reduced set, suggesting an improved latent structure. Nonetheless, some contamination in the anomaly islands remained, and to unveil the nature of these contaminants, we therefore color-coded each point with its RE score. Fig. \ref{fig: color-coded} shows that the points located near the anomaly islands tend to lie very close to the adopted threshold, the choice of which is ultimately arbitrary.

Although fewer than half of the starting features are the most discriminative features, the performance of the classifiers trained on the discriminative features (listed in Table \ref{tab:classification_results}) agrees with that of the classifiers trained on the whole feature set (listed in Table \ref{tab:classification_results} as well). All these results indicate that the discriminative features retain the essential information while providing advantages in terms of dimensionality reduction and interpretability. The comparison of our results with those by \cite{De_Cicco_2025} shows that this set of features does not share many similarities with the set obtained in that work. This is consistent with the fact that the features identified in \cite{De_Cicco_2025} are well suited for separating obscured AGN, whereas the features we obtained are particularly effective at distinguishing normal from anomalous obscured AGN. Furthermore, 7 out of the 12 most discriminative features shown in Fig. \ref{fig:feature_pdf} are in common with those identified in \cite{maruccia2025navigatingagnvariabilityselforganizing}, where the use of self-organizing maps revealed that in cells that are predominantly populated by AGN, these features significantly deviated from
the global mean of the same features in the remaining map. Taken together, these findings indicate that the set of features we obtained can capture the anomalous behavior.
\begin{figure}
    \centering
    \includegraphics[width=1.0\linewidth]{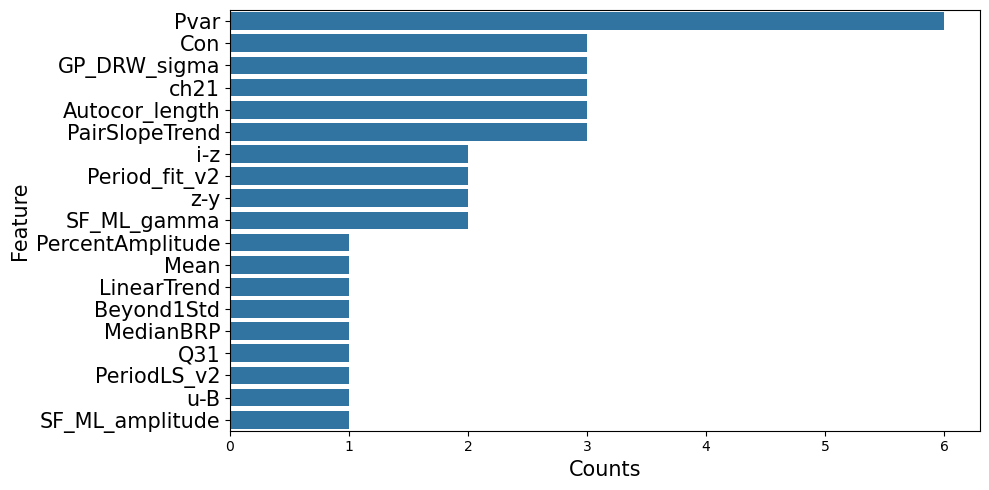}
    \caption{Bar plot of the most influential feature according to their SHAP value for the obscured AGN.}
    \label{fig:feature_tot}
\end{figure}

\begin{figure*}
    \centering
    \includegraphics[width=1.0\linewidth]{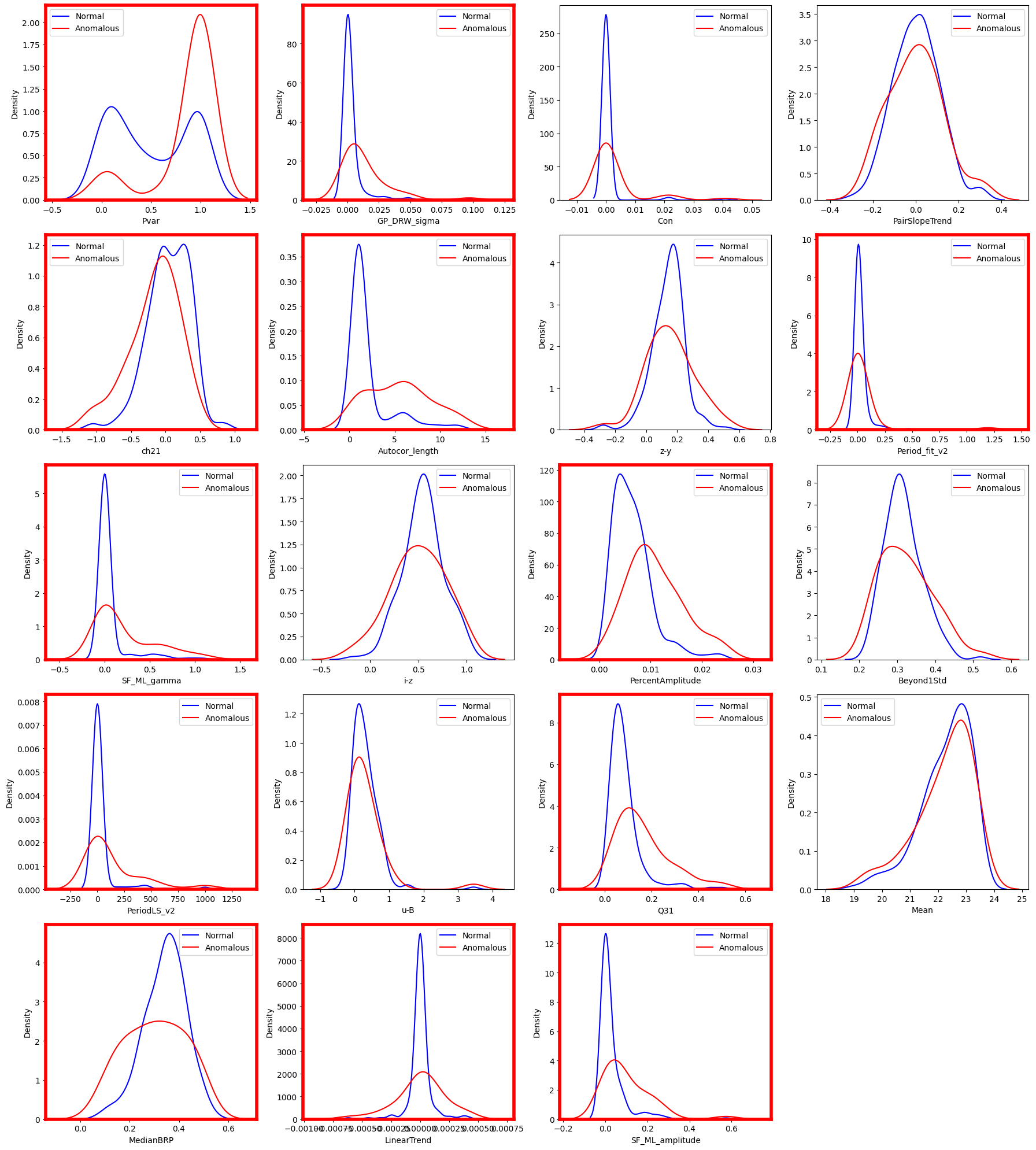}
    \caption{PDF for the 19 most influential features. The red lines represent the distribution for the normal obscured AGN, and the blue lines represent the anomalous AGN. The red frame highlights the statistically different distributions.}
    \label{fig:feature_pdf}
\end{figure*}

\begin{figure*}
\sidecaption
\includegraphics[width=12cm]{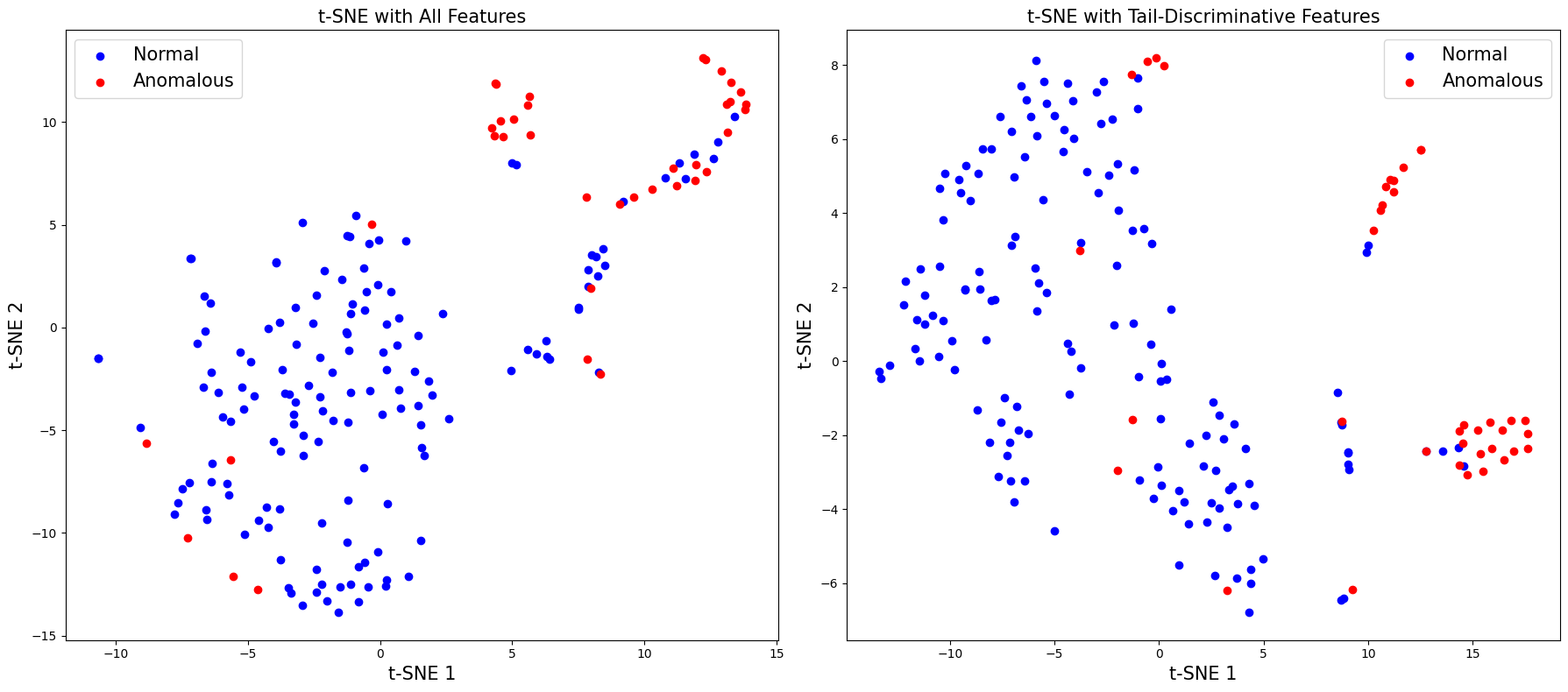}
\caption{Comparison of the t-SNE project of the whole feature set (\textit{left}) and the discriminative feature set for the obscured AGN (\textit{right}).}
\label{fig:comparison_tsne}
\end{figure*}
\begin{figure}
    \centering
    \includegraphics[width=1.0\linewidth]{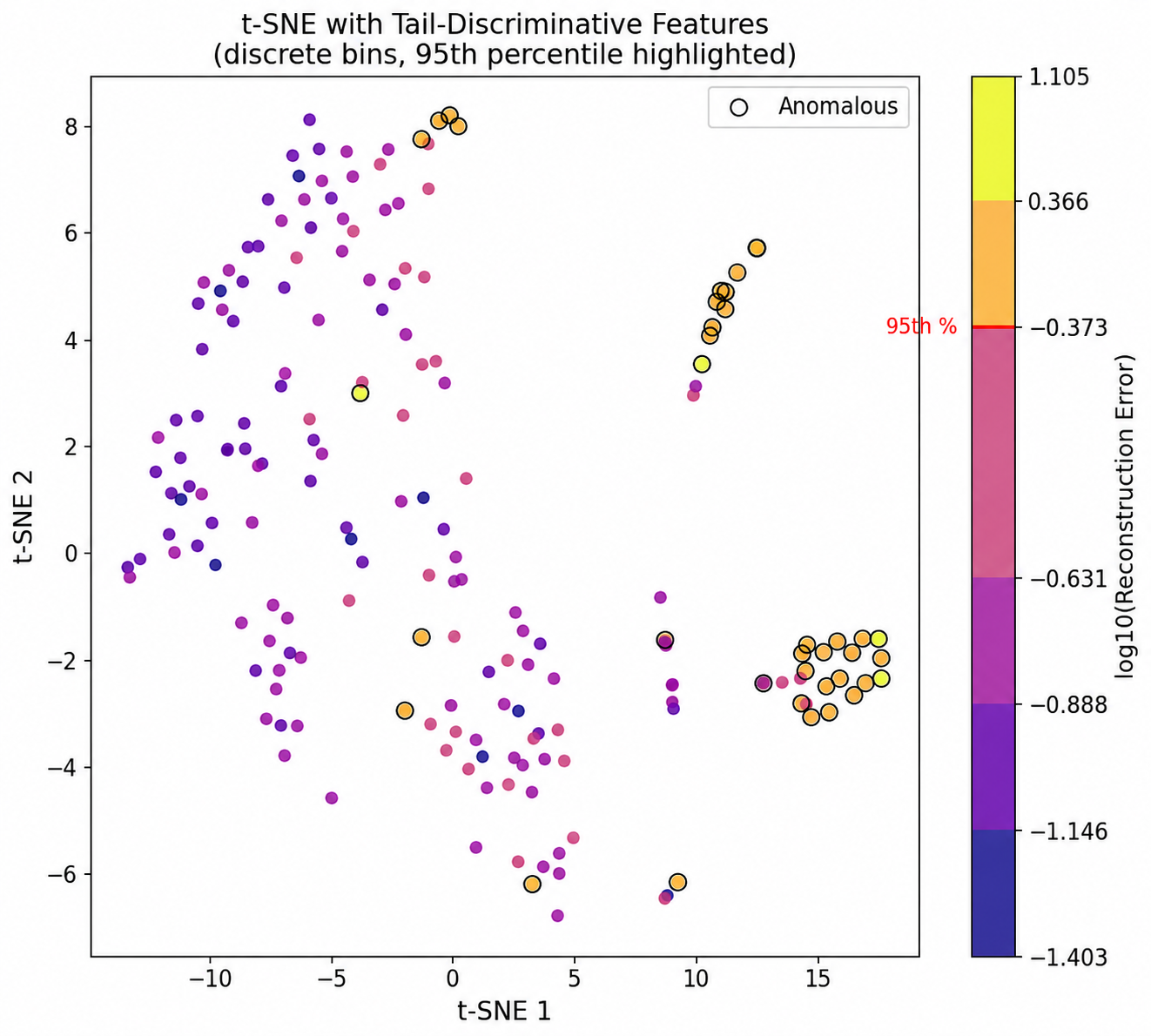}
    \caption{t-SNE representation of the obscured AGN population. The points are color-coded according to their RE score, and the black circles highlight the sources flagged as anomalous by the AE.}
    \label{fig: color-coded}
\end{figure}
\begin{table}
\caption{Cross-validated performance of random forest and gradient boosting using all vs. selected features.}
    \centering
    \renewcommand{\arraystretch}{1.5} 
    \resizebox{\linewidth}{!}{%
    \begin{tabular}{|l|c|c|c|}
    \hline
    \textbf{Model (Features)} & \textbf{ROC AUC} & \textbf{Precision} & \textbf{Recall} \\
    \hline
    \textbf{Random forest} (All) & $0.962 \pm 0.021$ & $0.852 \pm 0.128$ & $0.686 \pm 0.051$ \\
    \textbf{Gradient boosting} (All)  & $0.952 \pm 0.034$ & $0.828 \pm 0.103$ & $0.764 \pm 0.093$ \\
    \textbf{Random forest} (Selected) & $0.956 \pm 0.023$ & $0.843 \pm 0.091$ & $0.661 \pm 0.044$ \\
    \textbf{Gradient boosting} (Selected) & $0.943 \pm 0.028$ & $0.880 \pm 0.062$ & $0.736 \pm 0.081$ \\
    \hline
    \end{tabular}%
    }
    
    \label{tab:classification_results}
\end{table}

\subsection{Analysis of the  anomalous obscured AGN}
\label{sec:obscured}
As discussed previously, 14 of the 38 flagged anomalous AGN were classified as tail points, while the remaining 24 were not. A more detailed characterization of these flagged AGN can shed light on the features driving their flagging, whether it is physical or systematic. Moreover, by understanding the features that separate them, we can also gain clues about similar objects that were not identified as anomalies, thus refining our understanding of the boundaries defined by the autoencoder in feature space. The definition of all the features that are listed in the following is provided in Table \ref{tab:variability_features}.  We divided these 38 light curves into two primary categories based on the nature of their most contributing features: color-based features, and variability-based features.

Color-dominated anomalies: Six tail points and four non-tail points fall in this group; they are characterized by one of their colors as their most influential feature. These color features can reflect unusual spectral energy distributions (SEDs) compared to the typical obscured SEDs, or might simply indicate problematic photometry in one of the involved bands. Their placement in the color–color diagrams is shown in Figure \ref{fig:colcol_diagrams}. As shown in Table~\ref{tab:contour_counts}, tail points tend to occupy more extreme regions in color space than non-tail points, with a higher fraction lying outside the $1\sigma$ contour in all color planes. From a physical standpoint, the most informative color is u-B: extreme values indicate unusually heavy dust absorption along the line of sight, even within the obscured AGN population, suggesting that the tail points might represent the most heavily obscured sources in the sample.
\begin{table}
\centering
\small
\caption{Number of tail and non-tail points inside and outside the 1$\sigma$ contour of the obscured AGN distribution in each color-color diagram.}
\label{tab:contour_counts}
\begin{tabular}{lcccc}
\hline\hline
 & \multicolumn{2}{c}{Tail (14)} & \multicolumn{2}{c}{Non-tail (24)} \\
Color & In 1$\sigma$ & Out 1$\sigma$ & In 1$\sigma$ & Out 1$\sigma$ \\
\hline
$z-y$ vs $r-i$ & 8 (57.1\%) & 6 (42.9\%) & 15 (62.5\%) & 9 (37.5\%) \\
$i-z$ vs $r-i$ & 8 (57.1\%) & 6 (42.9\%) & 16 (66.7\%) & 8 (33.3\%) \\
$u-B$ vs $r-i$ & 8 (57.1\%) & 6 (42.9\%) & 19 (79.2\%) & 5 (20.8\%) \\
ch21 vs $r-i$  & 7 (50.0\%) & 7 (50.0\%) & 16 (66.7\%) & 8 (33.3\%) \\
\hline

\end{tabular}
\end{table}

Variability-dominated anomalies: The remaining 28 sources have variability features as their most influential feature. As summarized in Table~\ref{tab:variability_breakdown}, these can be broadly divided into three categories based on the physical interpretation of the driving feature. The first category includes 8 sources driven by \texttt{PercentAmplitude}, \texttt{Beyond1Std}, \texttt{Con}, \texttt{StetsonK}, or \texttt{SF\_ML\_gamma}, which might indicate a higher level of optical variability than expected for a purely obscured AGN, possibly suggesting an intermediate classification or instability in the accretion process \citep{Kawaguchi_1998, Hawkins_2002, Li_2018}. Notably, this category is evenly split between tail and non-tail points, suggesting that physically motivated anomalies can produce both strongly and moderately extreme feature values. The second category includes 6 sources driven by DRW-related features such as \texttt{GP\_DRW\_sigma} and \texttt{GP\_DRW\_tau}, where the anomalous detection is more likely a modeling artifact arising from the overestimation of short-timescale variability by the DRW model, particularly when the light curve is too sparsely sampled to constrain the relaxation time tau
reliably. The third category includes 14 sources driven by \texttt{Autocor\_length}, \texttt{PeriodLS\_v2}, or \texttt{Period\_fit\_v2}, which we interpret as systematic artifacts arising from poor light-curve sampling or seasonal gaps. For instance, \texttt{Autocor\_length} is sensitive to long-term correlations that can be artificially induced by an irregular cadence, while period-related features are prone to aliasing when the sampling is sparse. Interestingly, 12 out of 14 systematic sources are non-tail points, suggesting that systematic artifacts tend to produce moderate rather than strongly extreme anomalies, in contrast to physically motivated detections.
\begin{table}
\centering
\small
\caption{Breakdown of the variability-dominated anomalous obscured AGN by driver category and tail/non-tail classification.}
\label{tab:variability_breakdown}
\begin{tabular}{lccc}
\hline\hline
Category & Tail & Non-tail & Total \\
\hline
Physical$^a$   & 4 & 4  & 8  \\
Modeling$^b$   & 2 & 4  & 6  \\
Systematic$^c$ & 2 & 12 & 14 \\
\hline
Total          & 8 & 20 & 28 \\
\hline
\end{tabular}
\tablefoot{\tablefoottext{a}{Con, PercentAmplitude, Beyond1Std, StetsonK, and SF\_ML\_gamma.} \tablefoottext{b}{GP\_DRW\_sigma and GP\_DRW\_tau.} \tablefoottext{c}{Autocor\_length, PeriodLS\_v2, and Period\_fit\_v2.}}
\end{table}

The analysis presented above shows that tail points tend to be associated with more physically motivated anomalies, while non-tail points are more frequently driven by systematic or modeling artifacts. For this reason, we provide an in-depth characterization of the 14 tail-point anomalies below, focusing on their most influential feature and a possible physical explanation. They are referred to by their ID.

\paragraph{Color-dominated anomalies}

ID 9 For this AGN, the most influential feature is the color \textit{z}-\textit{y}, with a value of -0.27. However, this color is not the most extreme observed, as shown in Fig. \ref{fig:colcol_diagrams}, even though it is still outside the main distribution.
 
ID 29 The most influential feature for this source is the color \textit{i}-\textit{z}, with a value of -0.18. As in the case of ID 9, this point is not particularly extreme, and so it is difficult to address the anomaly detection using the color alone, meaning that other features contribute significantly to the detection of this anomaly. For example, this point is also in the tails of the \textit{u}-\textit{B} and \textit{B}-\textit{r} colors. This means that this anomalous behavior can be explained in general when the SED of this point is different from that of a normal obscured AGN.

ID 41 For this AGN, the most influential characteristic is the color \textit{u}-\textit{B}, with a value of 3.44, which is an abnormally high value, even for a heavily obscured AGN, and it is the highest value reached  by this color. The most reasonable explanation is that there must be a problem with the photometry in the \textit{u} band for this source.

ID 198 The feature contributing most is ch21, with a value of -1.07, and it is the largest one (in module). Fig. \ref{fig:colcol_diagrams} shows that this source clearly lies outside of the main distribution, highlighting its anomalous nature compared to the standard.

ID 260 Again, for this source, the most influential feature is the color ch21, with a value of -1.02. This source is clearly outside of the main distribution in the color-color diagram (Fig. \ref{fig:colcol_diagrams}).

ID 306 The color \textit{z}-\textit{y} is the most contributing feature for this source, with a value of 0.44. As in ID 9, this color lies outside of the main distribution, consistent with it being flagged as anomalous.

\paragraph{Variability-dominated anomalies}

ID 6 The most contributing feature for this AGN is \texttt{PercentAmplitude}, with a value of 0.02, the second highest value in the whole obscured sample. A high value of this feature can be an index of a higher level of variability, compared to the overall obscured population studied, possibly due to this AGN being part of some intermediate classification rather than being purely obscured.

ID 11 The most influential feature for this AGN is \texttt{Con}, with a value of 0.02, the second-highest value for this feature for the whole population. FATS computed this feature by counting the number of three consecutive data points that were brighter or fainter than $2\sigma$ and normalizing the number by $N-2$, meaning that this metric is sensitive to strong and persistent deviations in the light curve. A high value in this feature can be explained when the light curve has some epochs in which there is a huge deviation from the mean, which can be due to poor sampling. The light curve in Fig. \ref{fig:lc11} clearly shows an abrupt change in magnitude just before the first gap, most probably causing the detected anomaly.

ID 32 The most influential feature is \texttt{GP\_DRW\_$\sigma$}, with a value of 0.04. This is the second-highest value observed for this feature. An extreme value of this variable can be explained when the underlying model, the damped random walk (DRW; \citealt{2009ApJ...698..895K, 2010ApJ...721.1014M}), overestimates the short-term variability, namely on timescales shorter than the relaxation time $\tau$.

ID 191 The most influential feature for this source is the \texttt{Period\_fit\_v2}, with a value of 1.19, the highest in its distribution. A high value in this variable means that there is a high chance that the period obtained from the Lomb-Scargle \citep{1976Ap&SS..39..447L, 1982ApJ...263..835S} periodogram might be incorrect. In this case, the period is 0.09 days, which is far too short to be realistic since the minimum sampling is three days. This can be explained by considering the specific light curve in Fig. \ref{fig:lc191}, which is in fact almost flat.

ID 211 For this AGN, the most influential feature is again \texttt{GP\_DRW\_$\sigma$}, with a value of 0.09. The possible explanation is the same as for ID 32.

ID 219 The most influential feature is \texttt{Con}, with a value of 0.04, which is the highest in its distribution. The possible explanation is the same as for ID 11.

ID 340 The most influential feature for this source is \texttt{PeriodLS\_v2}, with a value of 999.99 days, which is near the overall coverage of the whole light curve (three years), and it is one of the most extreme values observed for this parameter in the dataset. A long period like this might result from poor sampling in the light curve. Consequently, the anomalous classification and the inflated period estimate might be driven by the same underlying issue in the temporal coverage.

ID 380 \texttt{Beyond1std} is the most strongly contributing feature for this source, with a value of 0.51, and it is the highest value observed for this variable. Given its definition, this high value means that more than half of the points lie outside $1\sigma$ from the mean. This type of behavior is not typical for obscured AGN, which are expected to exhibit modest variation in their light curves, again suggesting an intermediate classification and not that this source is a purely obscured AGN.

A coherent picture emerges from this analysis. Of the 38 anomalous obscured AGN, 10 are color dominated, suggesting unusual SEDs or photometric issues, while 28 are variability dominated. Of the latter, 8 sources show physically motivated variability, 6 are likely modeling artifacts, and 14 are consistent with systematic effects from poor sampling or seasonal gaps. Crucially, the tail/non-tail classification proves to be a useful proxy for the physical versus systematic origin of the anomaly: tail points are predominantly associated with color anomalies and physically motivated variability, while systematic and modeling artifacts are more frequently found among non-tail points.

Furthermore, we gain valuable insight into how the AE detects anomalous behaviors within the obscured AGN population. It is essential to identify the specific feature, or combination of features, that drives an anomaly because it enables us to interpret whether the source detection as an anomaly arises from physical properties, statistical outliers, or systematic effects. This information is important for several reasons.

First, it allows us to remove certain candidates from further investigations. Second, by recognizing common patterns among anomalous sources, we can better understand how the AE defines the boundaries between normal and anomalous points in the learned feature space. Finally, this approach emphasizes the importance of incorporating diverse types of features, such as variability, photometric, and morphological features, since anomalous behavior may arise from any of those.

\begin{figure*}
    \sidecaption
    \includegraphics[width=12cm]{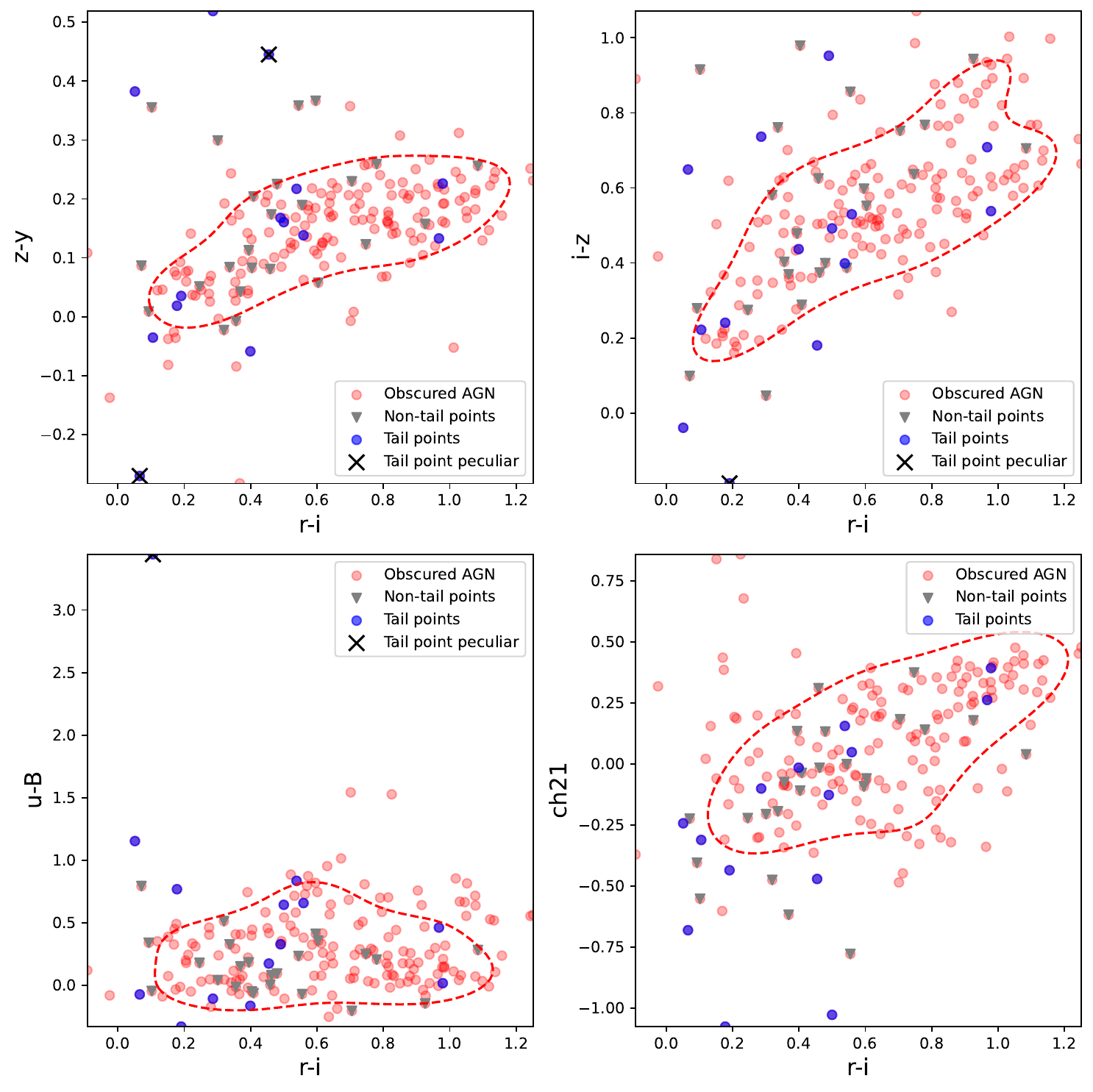}
    \caption{Color-color diagrams for the obscured AGN population. The red points are all the obscured AGN, the blue points represent the tail points defined in the text, the gray triangles represent the non-tail points also defined in the text, and the black crosses represent tail points for which the color on the corresponding y-axis is the most influential feature. Last, we highlight the $1\sigma$ threshold in each plot (dashed black line).}
    \label{fig:colcol_diagrams}
\end{figure*}

\section{Summary and conclusions}
\label{conclusion}
We have explored the use of autoencoders in the field of anomaly detection of AGN. We made predictions on the LS after the training phase on the US, obtaining a percentage of outliers of $11.18\%$. Specifically, 272 sources were classified as anomalous, of which 242 are AGN, 16 are galaxies, and 14 are stars. The flagging rate varies strongly in the AGN subclasses: unobscured AGN are flagged at $90.7\%$, obscured AGN at $31.1\%$, and unclassified AGN at only $1.5\%$. This is consistent with the expected differences in optical variability between these populations.

To assess whether the AE is sensitive to AGN-specific properties and does not act as a generic outlier detector, we extended the SHAP analysis to the full LS. The results show that anomalous AGN are primarily flagged due to variability features such as \texttt{ExcessVar}, \texttt{GP\_DRW\_sigma}, and \texttt{Meanvariance}, which are directly linked to accretion disk variability. In contrast, anomalous galaxies and stars are dominated by the structure function slope \texttt{SF\_ML\_gamma}, pointing to a fundamentally different origin. Furthermore, only 7 out of 15 top features are shared between anomalous and normal AGN, confirming that flagged AGN are genuine outliers within their class.

We then concentrated on the analysis of the 38 anomalous obscured AGN. To characterize these sources, we used the SHAP framework to understand which features drove their anomalous detection. From the SHAP analysis, combined with the KS test, we identified a sub-feature set of 12 features that are most influential for separating normal from anomalous obscured AGN. The performance of classifiers trained on this reduced feature set was comparable to that obtained using the full feature set, confirming that the selected features retain the essential discriminative information while providing a more interpretable and computationally efficient alternative.

From an in-depth SHAP analysis of the obscured AGN population, we obtained that out of the 38 anomalous obscured AGN, 10 are color dominated, suggesting unusual SED or photometric issues, while 28 are variability dominated. Of the latter, 8 sources show physically motivated variability consistent with intermediate AGN classifications or accretion instabilities, 6 are likely modeling artifacts from DRW overestimation, and 14 are consistent with systematic effects from poor sampling or seasonal gaps. Crucially, the tail/non-tail classification proves to be a useful proxy for the physical versus systematic origin: tail points are predominantly associated with physically motivated anomalies, while systematic and modeling artifacts are more frequently found among non-tail points. 

Beyond these specific results, our study demonstrated that feature-driven autoencoders can provide a reliable and interpretable framework for anomaly detection in AGN light curves. These findings have broader implications for upcoming surveys, such as the LSST, where unsupervised models will be essential for identifying rare or unexpected sources among billions of observations. Overall, our results confirmed that combining feature-based representations with explainable ML tools can significantly enhance the interpretability and scientific return of automated anomaly detection in astrophysical time-series data.

\begin{acknowledgements}
This paper is supported by Italian Research Center on High Performance Computing Big Data and Quantum Computing (ICSC), project funded by European Union - NextGenerationEU - and National Recovery and Resilience Plan (NRRP) - Mission 4 Component 2 within the activities of Spoke 3 (Astrophysics and Cosmos Observations). DD acknowledges PON R\&I 2021, CUP E65F21002880003, Fondi di Ricerca di Ateneo (FRA), linea C, progetto TORNADO, and the financial contribution from PRIN-MIUR 2022 and from the Timedomes grant within the ``INAF 2023 Finanziamento della Ricerca Fondamentale. A.B.K. and D.I. acknowledge funding provided by the University of Belgrade - Faculty of Mathematics (the contract 451-03-33/2026-03/200104) through the grant by the Ministry of Science, Technological Development and Innovation of the Republic of Serbia. The research leading to these results has received funding from the EU HORIZON-MSCA-2023-DN Project 101168906 "TALES: Time-domain Analysis to study the Life-cycle and Evolution of Supermassive black holes."
\end{acknowledgements}

\bibliographystyle{aa}
\bibliography{biblio} 

\begin{appendix}

\twocolumn[
\begin{minipage}{\columnwidth}
\section{Feature set and distributions}
Here, we describe the features used for this work and show their distribution. Figure~\ref{fig:fetaure_comparison} shows the feature distribution, while Table \ref{tab:variability_features} lists the features themselves.
\end{minipage}

\vspace{1em}

{\centering
\resizebox{0.75\textwidth}{!}{\includegraphics{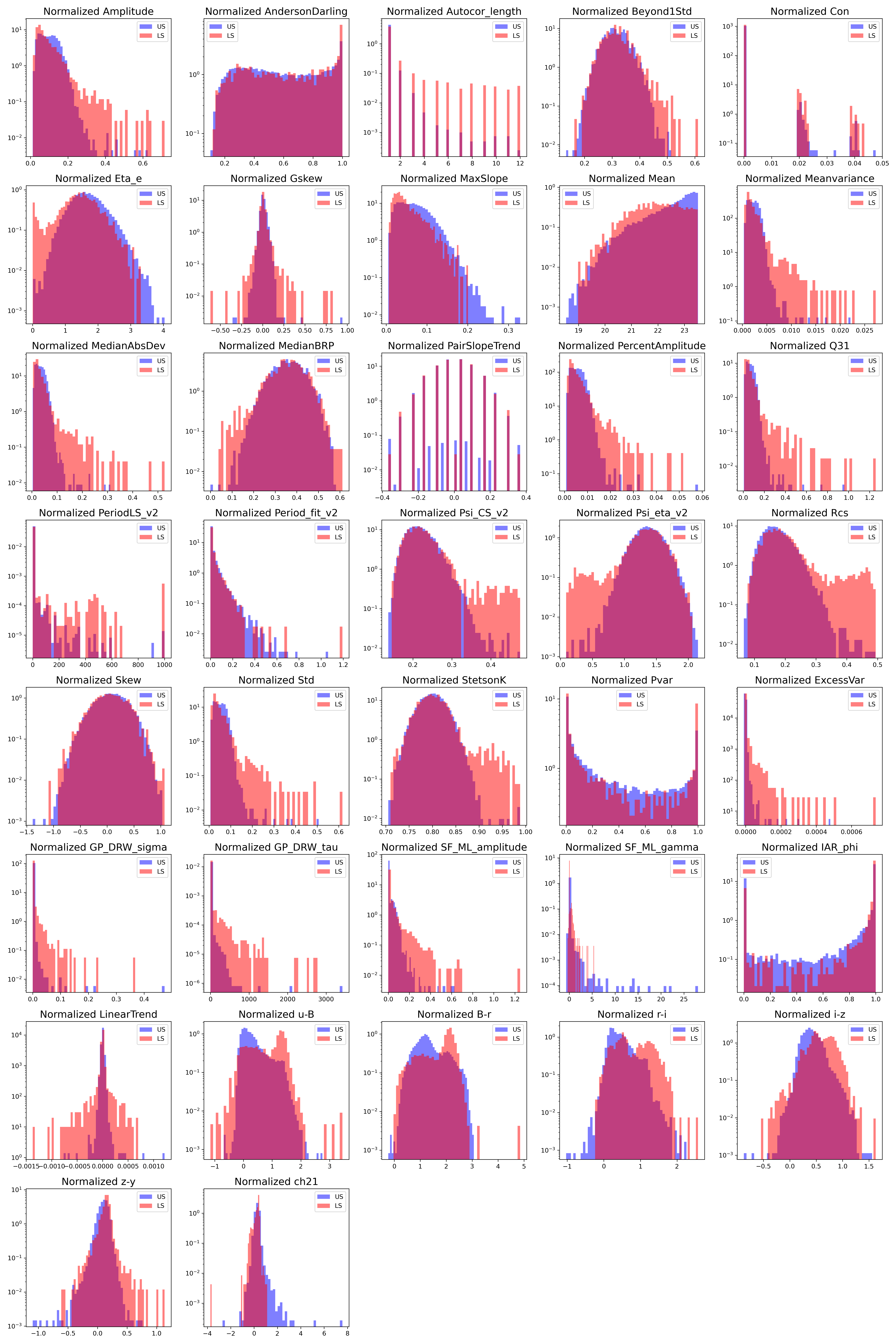}}\par
\refstepcounter{figure}
\textbf{Fig.~\thefigure.} Normalized feature distributions for the LS and US. The distributions are shown as probability densities, allowing a direct comparison between the two samples despite their different sizes. The LS is broadly representative of the US across all features.
\label{fig:fetaure_comparison}\par}
\vspace{1em}]

\begin{table*}
    \centering
    \caption{Features we used.}
    \resizebox{\linewidth}{!}{
    \begin{tabular}{|p{3.5cm}|p{10.5cm}|p{2.0cm}|p{4.0cm}|}
        \hline
        \textbf{Feature} & \textbf{Description} & \textbf{Units} & \textbf{Reference} \\
        \hline
        \multicolumn{4}{|c|}{\textbf{Variability features}} \\
        \hline
        SF\_ML\_amplitude & RMS magnitude difference of the Structure Function (SF), computed over a 1 yr timescale & mag & \cite{Schmidt_2010} \\
        SF\_ML\_gamma & Logarithmic gradient of the mean change in magnitude & -- & \cite{Schmidt_2010} \\
        GP\_DRW\_tau & Relaxation time (i.e., time necessary for the time series to become uncorrelated), from a Damped Random Walk (DRW) model & days & \cite{Graham_2017} \\
        GP\_DRW\_sigma & Variability amplitude of the time series at short timescales ($t \ll \tau$), from a DRW model & mag & \cite{Graham_2017} \\
        ExcessVar & Measure of the intrinsic variability amplitude, defined as $\sigma^2_{\rm rms} = (S^2 - \overline{\sigma^2_{\rm err}})/\overline{x}^2$, where $S^2$ is the sample variance, $\overline{\sigma^2_{\rm err}}$ the mean squared error, and $\overline{x}$ the mean flux & -- & \cite{Allevato_2013} \\
        Pvar & Probability that the source is intrinsically variable & -- & \cite{1996ApJ...473..763M} \\
        IAR\_phi & Level of autocorrelation using a discrete-time representation of a DRW model & -- & \cite{Eyheramendy_2018} \\
        Mean & Mean magnitude & mag & \cite{nun2015fatsfeatureanalysistime} \\
        Con & Fraction of three consecutive data points deviating by more than $2\sigma$ from the mean, normalized by $N-2$; sensitive to strong and persistent deviations in the light curve & -- & \cite{Kim_2011} \\
        Amplitude & Half of the difference between the median of the maximum 5\% and of the minimum 5\% magnitudes & mag & \cite{Richards_2011} \\
        AndersonDarling & Statistical test of whether the light curve magnitude distribution follows a Gaussian; higher values indicate stronger departure from normality & -- & \cite{nun2015fatsfeatureanalysistime} \\
        Autocor\_length & Lag value (in number of observations) where the autocorrelation function first drops below $\eta^e$; large values indicate long-term correlated variability & obs & \cite{Kim_2011} \\
        Beyond1Std & Fraction of photometric points lying beyond $1\sigma$ from the mean magnitude & -- & \cite{Richards_2011} \\
        Eta\_e & Ratio of the mean of the squares of successive magnitude differences to the variance of the light curve & -- & \cite{Kim_2014} \\
        Gskew & Median-based skewness measure: $(m_{31} + m_{69} - 2\,m_{50})/(m_{31} - m_{69})$, where $m_p$ is the $p$-th percentile of the magnitude distribution & -- & -- \\
        LinearTrend & Slope of a linear fit to the light curve & mag/day & \cite{Richards_2011} \\
        MaxSlope & Maximum absolute magnitude slope between two consecutive observations & mag/day & \cite{Richards_2011} \\
        Meanvariance & Ratio of the standard deviation to the mean magnitude & -- & \cite{nun2015fatsfeatureanalysistime} \\
        MedianAbsDev & Median of the absolute deviations from the median magnitude & mag & \cite{Richards_2011} \\
        MedianBRP & Fraction of photometric points within amplitude/10 of the median magnitude & -- & \cite{Richards_2011} \\
        PeriodLS\_v2 & Period obtained from the Lomb-Scargle periodogram using the P4J Python package (\url{https://github.com/phuijse/P4J}) & days & \cite{Huijse_2018} \\
        PairSlopeTrend & Fraction of increasing first differences minus decreasing ones over the last 30 time-sorted magnitude measures & -- & \cite{Richards_2011} \\
        PercentAmplitude & Largest percentage difference between either the maximum or minimum magnitude and the median magnitude & -- & \cite{Richards_2011} \\
        Q31 & Difference between the third and the first quartile of the magnitude distribution & mag & \cite{Kim_2014} \\
        Period\_fit\_v2 & False-alarm probability of the largest Lomb-Scargle periodogram peak; values near 1 indicate an unreliable period estimate & -- & \cite{Kim_2011} \\
        Psi\_CS\_v2 & Range of a cumulative sum applied to the phase-folded light curve & mag & \cite{Kim_2011} \\
        Psi\_eta\_v2 & Eta\_e index calculated from the phase-folded light curve & -- & \cite{Kim_2014} \\
        Rcs & Range of a cumulative sum of the light curve & mag & \cite{Kim_2011} \\
        Skew & Skewness of the magnitude distribution & -- & \cite{Kim_2011} \\
        Std & Standard deviation of the light curve & mag & \cite{nun2015fatsfeatureanalysistime} \\
        StetsonK & Robust kurtosis measure of the light curve & -- & \cite{Kim_2011} \\
        \hline
        \multicolumn{4}{|c|}{\textbf{Morphological feature}} \\
        \hline
        class\_star & HST stellarity index (0 = extended source, 1 = point source) & -- & \cite{Koekemoer_2007, Scoville_2007} \\
        \hline
        \multicolumn{4}{|c|}{\textbf{Photometric features}} \\
        \hline
        u-B & CFHT $u$ magnitude -- Subaru $B$ magnitude & mag & \cite{Laigle_2016} \\
        B-r & Subaru Suprime-Cam $B$ mag -- Subaru Suprime-Cam $r+$ mag & mag & \cite{Laigle_2016} \\
        r-i & Subaru Suprime-Cam $r+$ mag -- Subaru Suprime-Cam $i+$ mag & mag & \cite{Laigle_2016} \\
        i-z & Subaru Suprime-Cam $i+$ mag -- Subaru Suprime-Cam $z++$ mag & mag & \cite{Laigle_2016} \\
        z-y & Subaru Suprime-Cam $z++$ mag -- Subaru Hyper-Suprime-Cam $y$ mag & mag & \cite{Laigle_2016} \\
        ch21 & \textit{Spitzer} 4.5$\mu$m (channel2) -- 3.6$\mu$m (channel1) & mag & \cite{Laigle_2016} \\
        \hline
    \end{tabular}}
    \tablefoot{Units of `--' indicate dimensionless quantities.}
    \label{tab:variability_features}
\end{table*}

\clearpage
\section{Obscured AGN light curves}
Here, we show the anomalous light curves of the peculiar obscured AGN discussed in the text to better highlight why these sources are classified as anomalous by the AE. In all the following figures, the blue solid line highlights the mean magnitude, and the two dashed blue lines are 2-sigma from the mean.

\begin{figure}[H]
    \centering
    \includegraphics[width=1.0\linewidth]{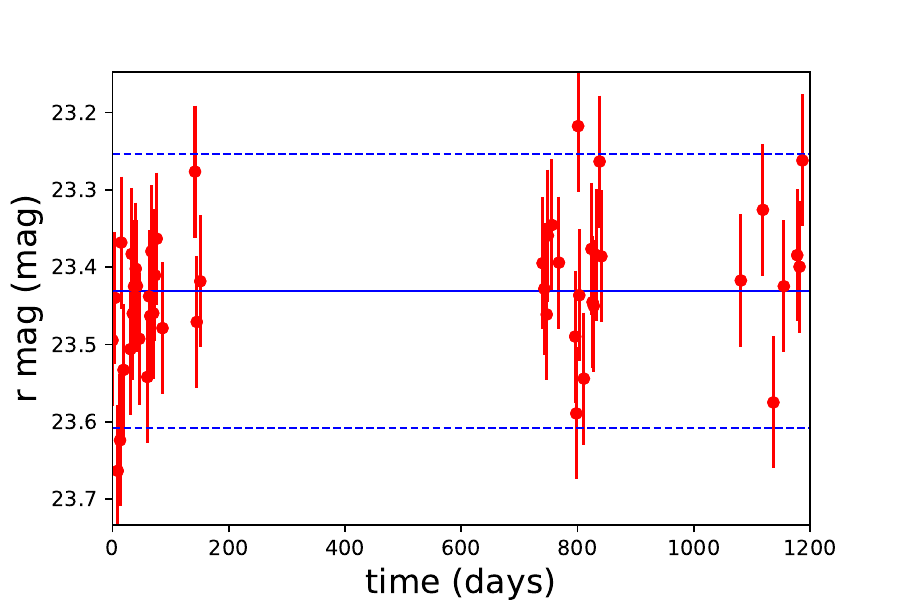}
    \caption{Light curve of ID 9. The most influential feature is color z-y.}
    \label{fig:lc9}
\end{figure}

\begin{figure}[H]
    \centering
    \includegraphics[width=1.0\linewidth]{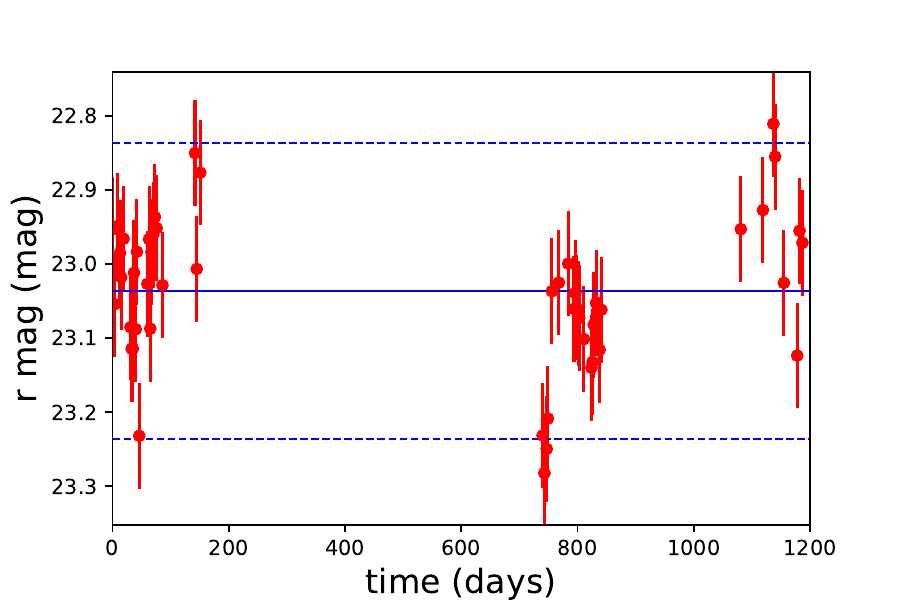}
    \caption{Light curve of ID 29. The most influential feature is color i-z.}
    \label{fig:lc29}
\end{figure}

\begin{figure}[H]
    \centering
    \includegraphics[width=1.0\linewidth]{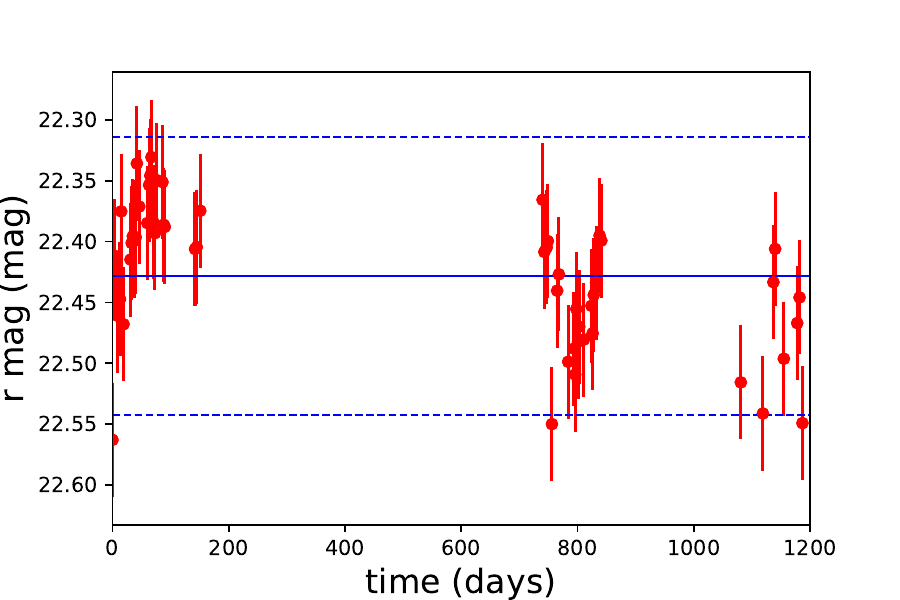}
    \caption{Light curve of ID 41. The most influential feature is color u-B.}
    \label{fig:lc41}
\end{figure}

\begin{figure}[H]
    \centering
    \includegraphics[width=1.0\linewidth]{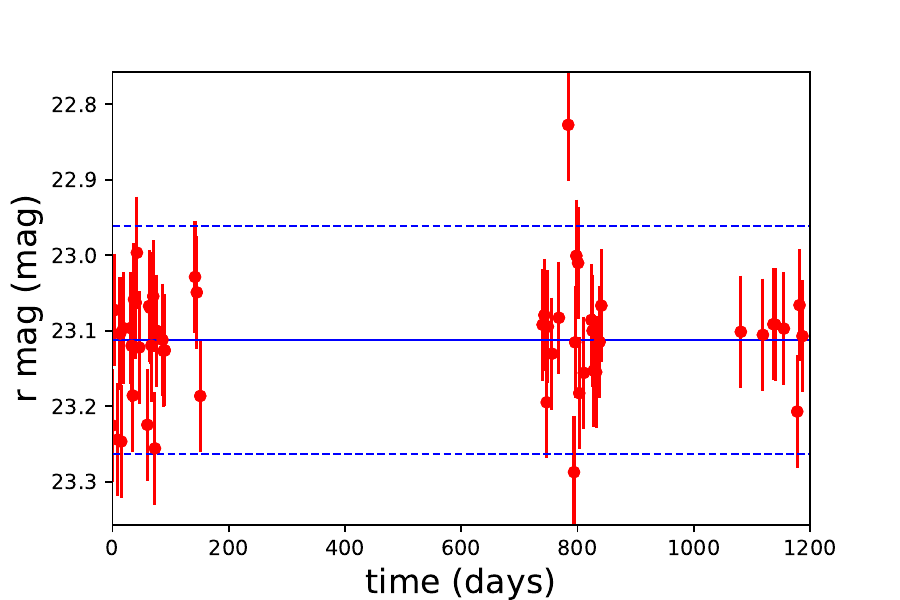}
    \caption{Light curve of ID 198. The most influential feature is ch21.}
    \label{fig:lc198}
\end{figure}

\begin{figure}[H]
    \centering
    \includegraphics[width=1.0\linewidth]{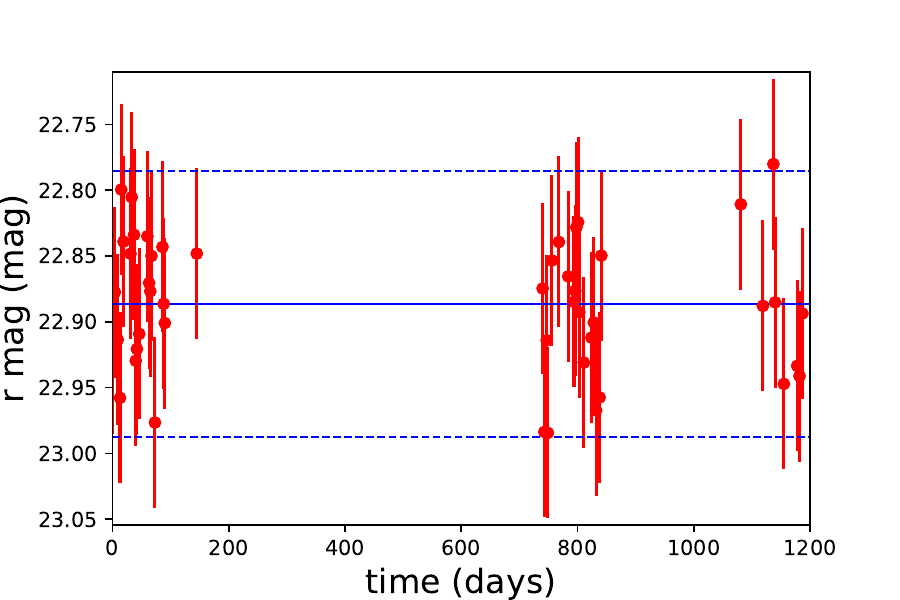}
    \caption{Light curve of ID 260. The most influential feature is ch21.}
    \label{fig:lc260}
\end{figure}

\begin{figure}[H]
    \centering
    \includegraphics[width=1.0\linewidth]{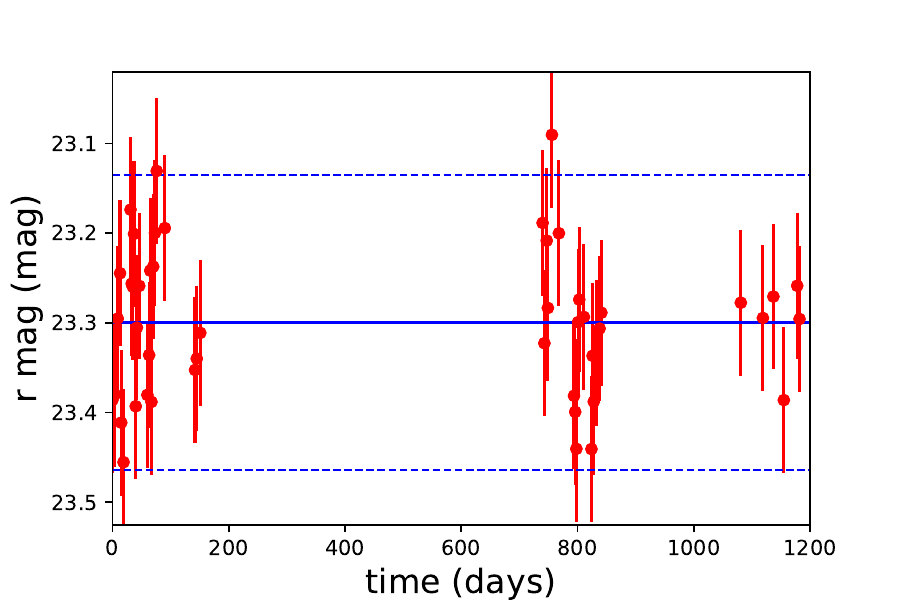}
    \caption{Light curve of ID 306. The most influential feature is color z-y.}
    \label{fig:lc306}
\end{figure}

\begin{figure}[H]
    \centering
    \includegraphics[width=1.0\linewidth]{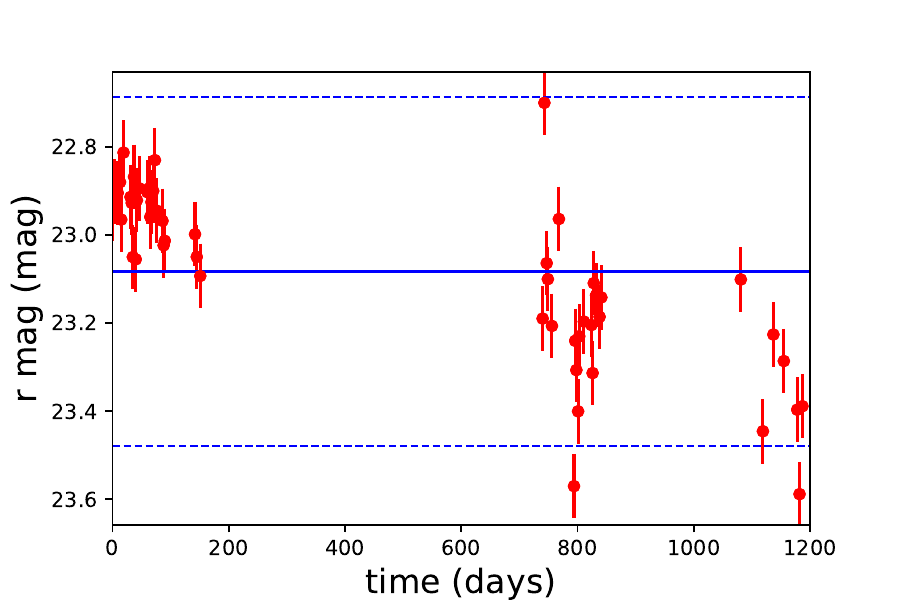}
    \caption{Light curve of ID 6. The most influential feature is \texttt{PercentAmplitude}.}
    \label{fig:lc6}
\end{figure}

\begin{figure}[H]
    \centering
    \includegraphics[width=1.0\linewidth]{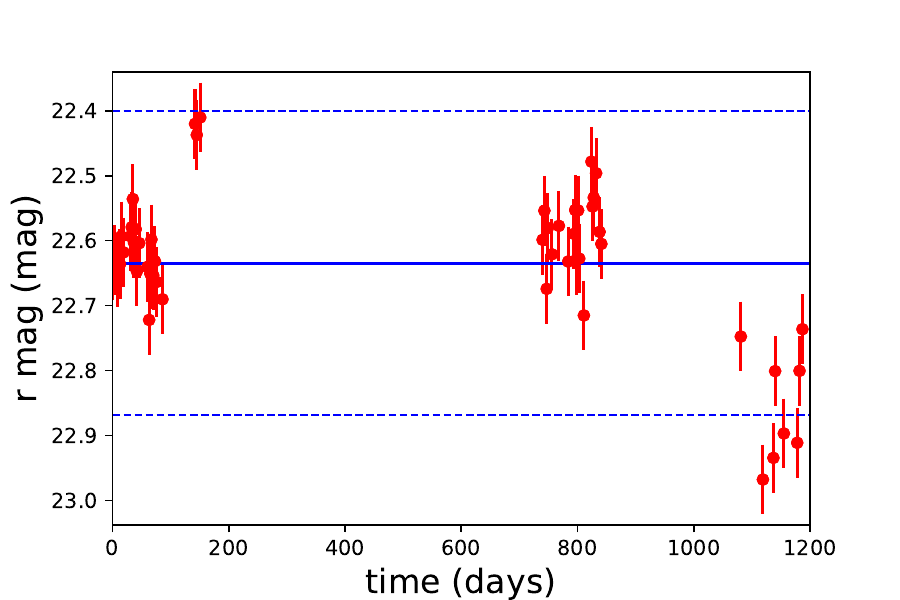}
    \caption{Light curve of ID 11. The most influential feature is \texttt{Con}.}
    \label{fig:lc11}
\end{figure}

\begin{figure}[H]
    \centering
    \includegraphics[width=1.0\linewidth]{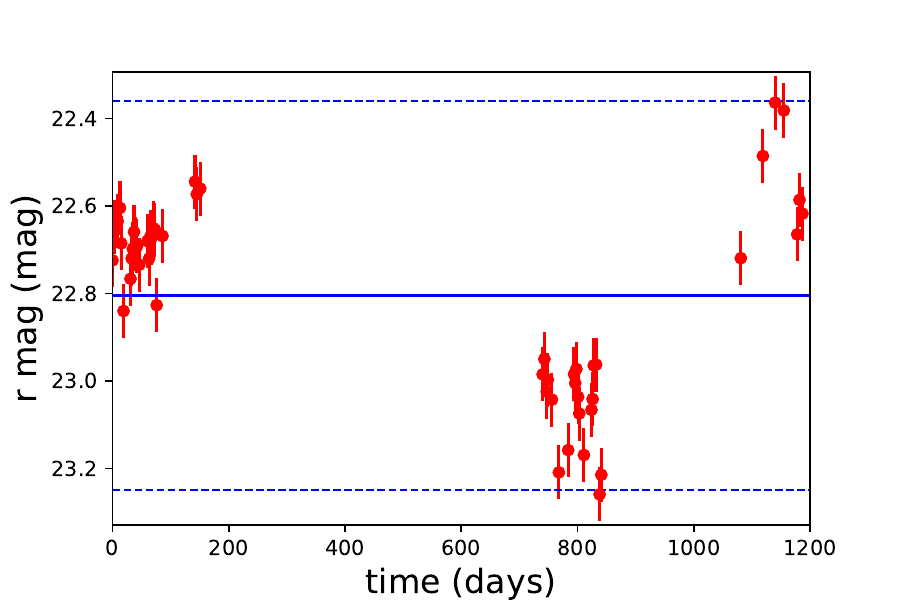}
    \caption{Light curve of ID 32. The most influential feature is \texttt{GP\_DRW\_$\sigma$}.}
    \label{fig:lc32}
\end{figure}

\begin{figure}[H]
    \centering
    \includegraphics[width=1.0\linewidth]{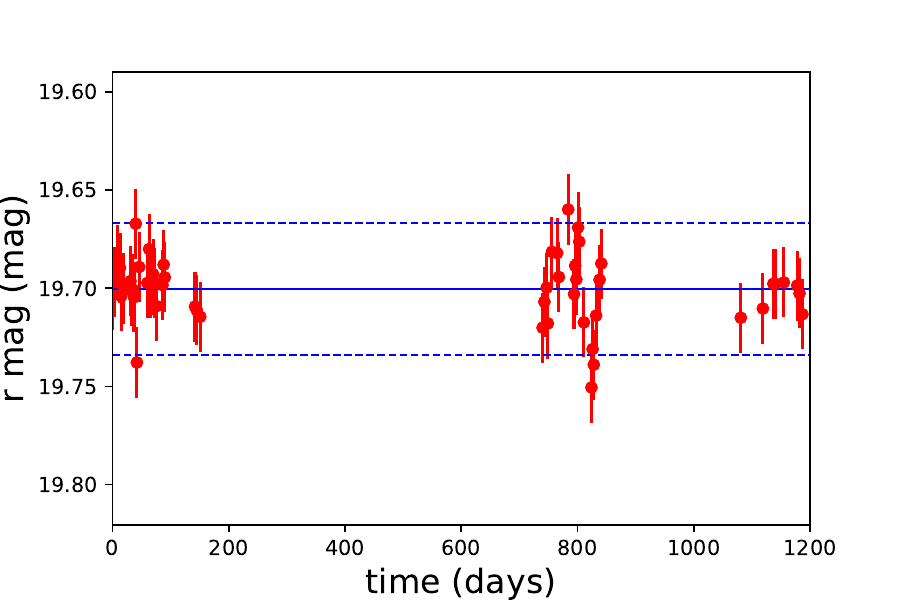}
    \caption{Light curve of ID 191. The most influential feature is \texttt{Period\_fit\_v2}.}
    \label{fig:lc191}
\end{figure}

\begin{figure}[H]
    \centering
    \includegraphics[width=1.0\linewidth]{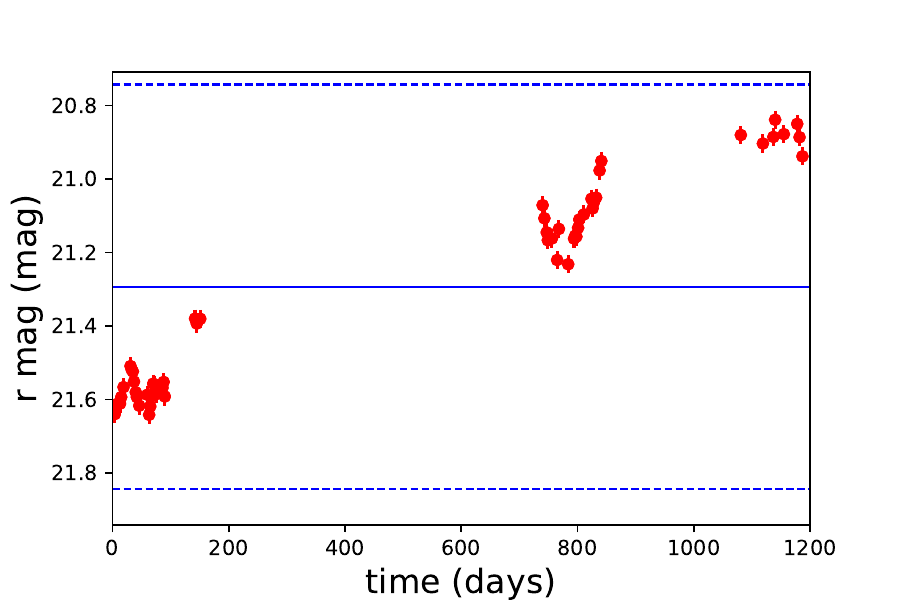}
    \caption{Light curve of ID 211. The most influential feature is \texttt{GP\_DRW\_$\sigma$}.}
    \label{fig:lc211}
\end{figure}

\begin{figure}[H]
    \centering
    \includegraphics[width=1.0\linewidth]{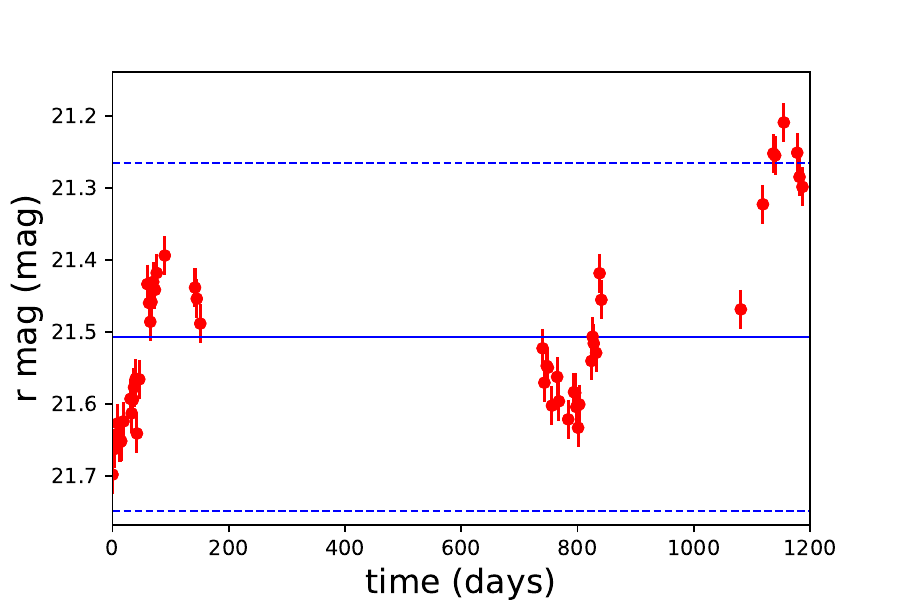}
    \caption{Light curve of ID 219. The most influential feature is \texttt{Con}.}
    \label{fig:lc219}
\end{figure}

\begin{figure}[H]
    \centering
    \includegraphics[width=1.0\linewidth]{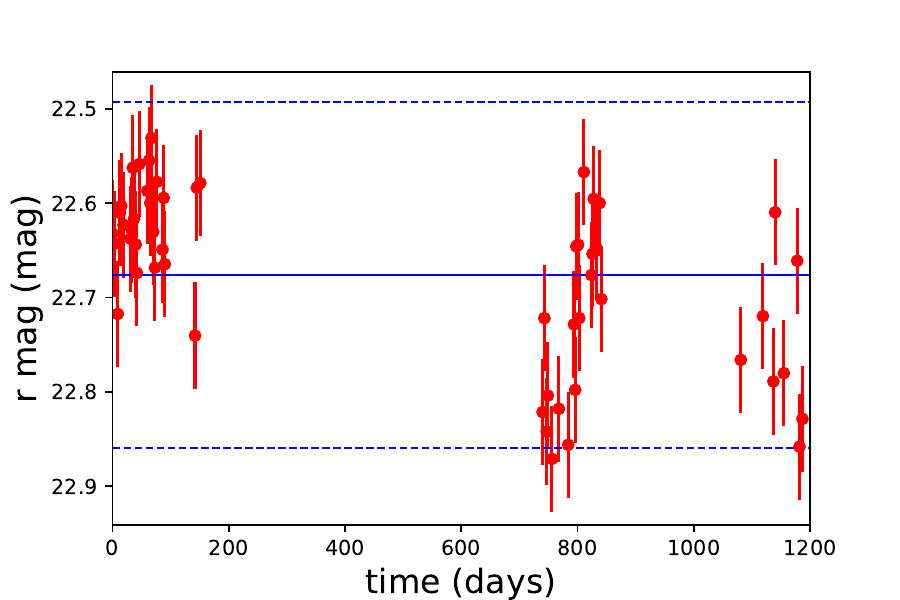}
    \caption{Light curve of ID 340. The most influential feature is \texttt{PeriodLS\_v2}.}
    \label{fig:lc340}
\end{figure}

\begin{figure}[H]
    \centering
    \includegraphics[width=1.0\linewidth]{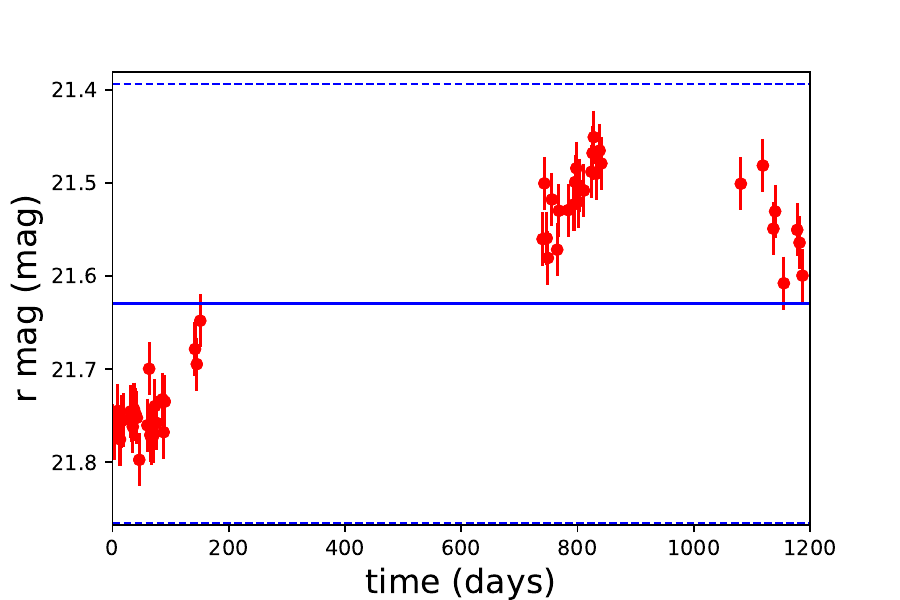}
    \caption{Light curve of ID 380. The most influential feature is \texttt{Beyond1Std}.}
    \label{fig:lc380}
\end{figure}

\section{Stacked images of the 14 anomalies}
Stacked images of the most interesting obscured AGN.

\begin{figure}[H]
    \centering
    \includegraphics[width=1.0\linewidth]{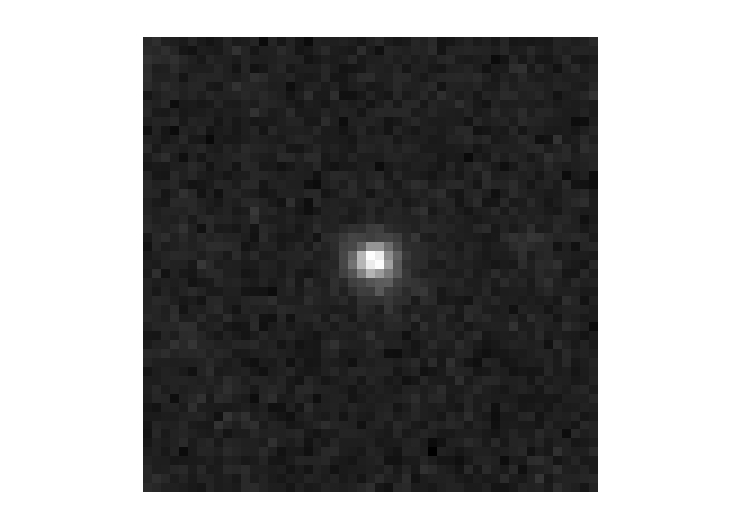}
    \caption{Stacked image of ID 9. The most influential feature is color z-y.}
    \label{fig:anomaly_9}
\end{figure}

\begin{figure}[H]
    \centering
    \includegraphics[width=1.0\linewidth]{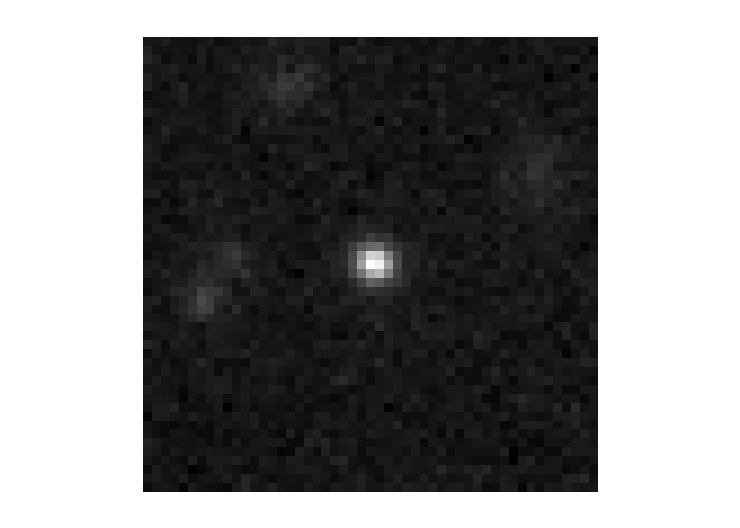}
    \caption{Stacked image of ID 29. The most influential feature is color i-z.}
    \label{fig:anomaly_29}
\end{figure}

\begin{figure}[H]
    \centering
    \includegraphics[width=1.0\linewidth]{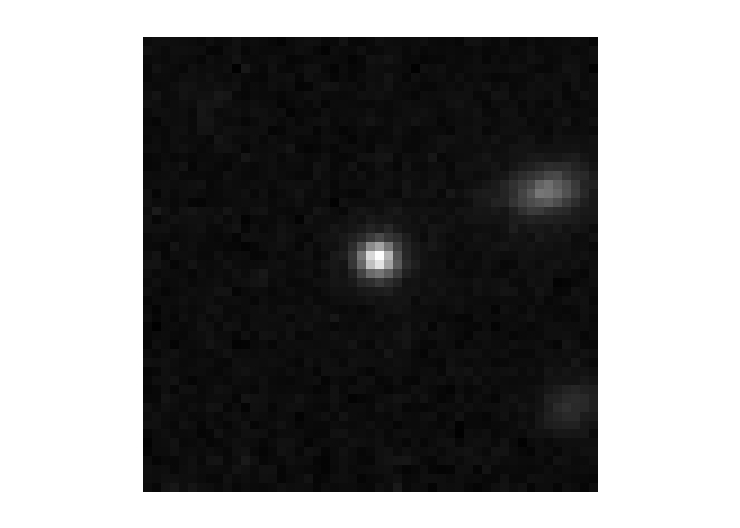}
    \caption{Stacked image of ID 41. The most influential feature is color u-B.}
    \label{fig:anomaly_41}
\end{figure}

\begin{figure}[H]
    \centering
    \includegraphics[width=1.0\linewidth]{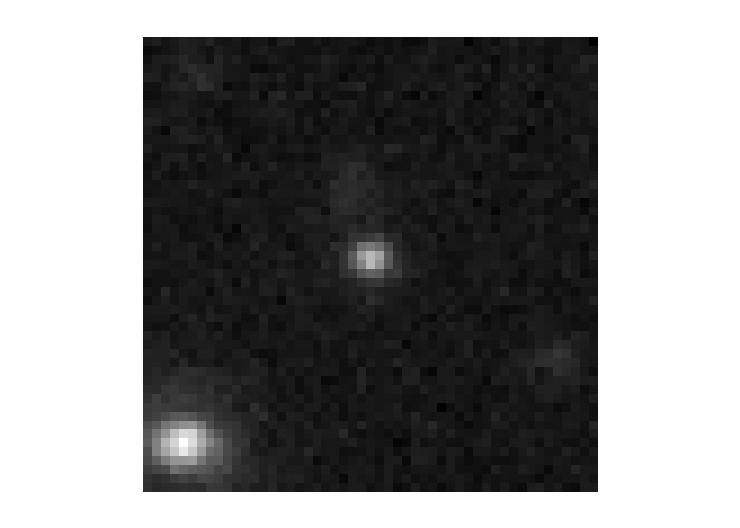}
    \caption{Stacked image of ID 198. The most influential feature is ch21.}
    \label{fig:anomaly_198}
\end{figure}

\begin{figure}[H]
    \centering
    \includegraphics[width=1.0\linewidth]{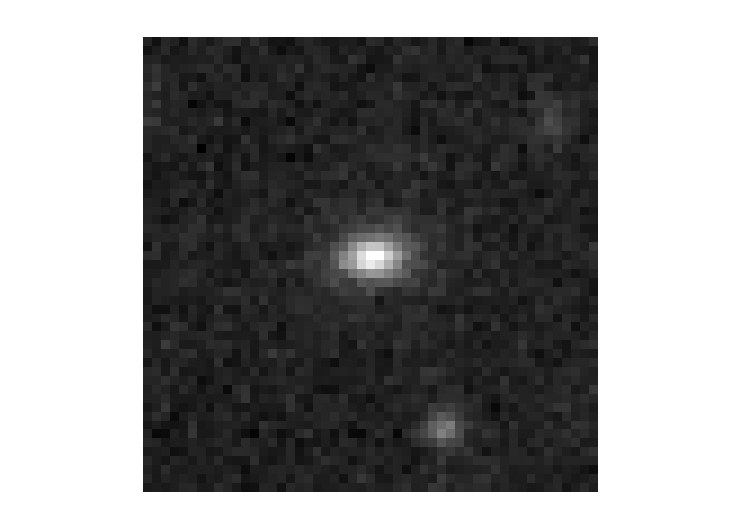}
    \caption{Stacked image of ID 260. The most influential feature is ch21.}
    \label{fig:anomaly_260}
\end{figure}

\begin{figure}[H]
    \centering
    \includegraphics[width=1.0\linewidth]{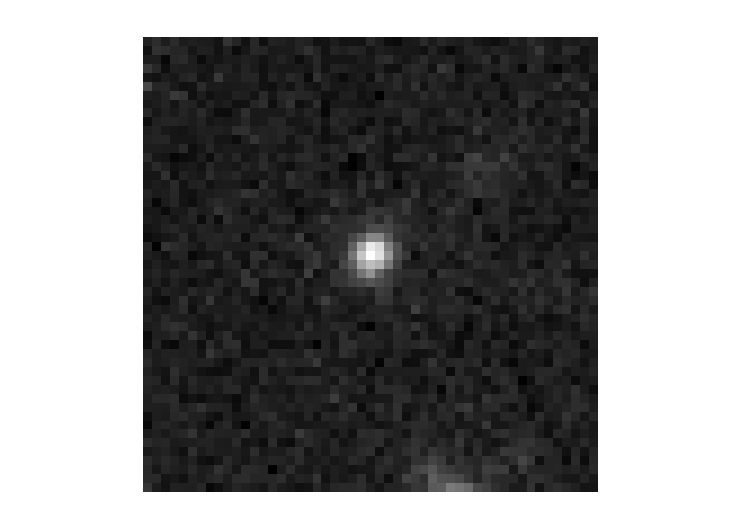}
    \caption{Stacked image of ID 306. The most influential feature is color z-y.}
    \label{fig:anomaly_306}
\end{figure}

\begin{figure}[H]
    \centering
    \includegraphics[width=1.0\linewidth]{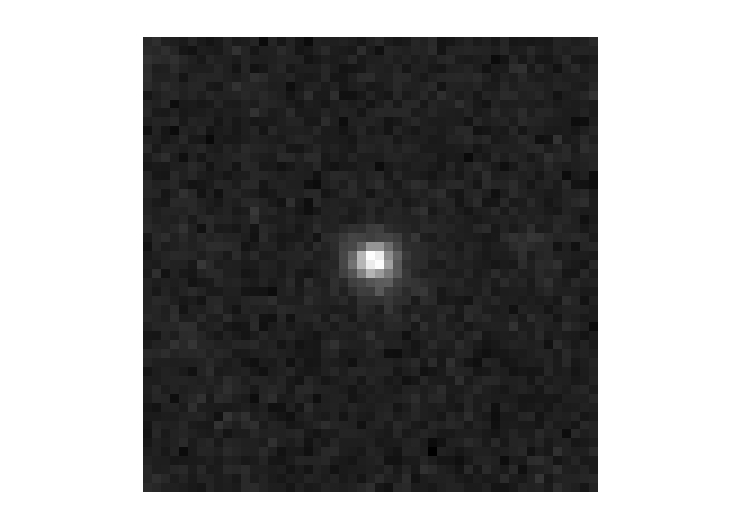}
    \caption{Stacked image of ID 6. The most influential feature is \texttt{PercentAmplitude}.}
    \label{fig:anomaly_6}
\end{figure}

\begin{figure}[H]
    \centering
    \includegraphics[width=1.0\linewidth]{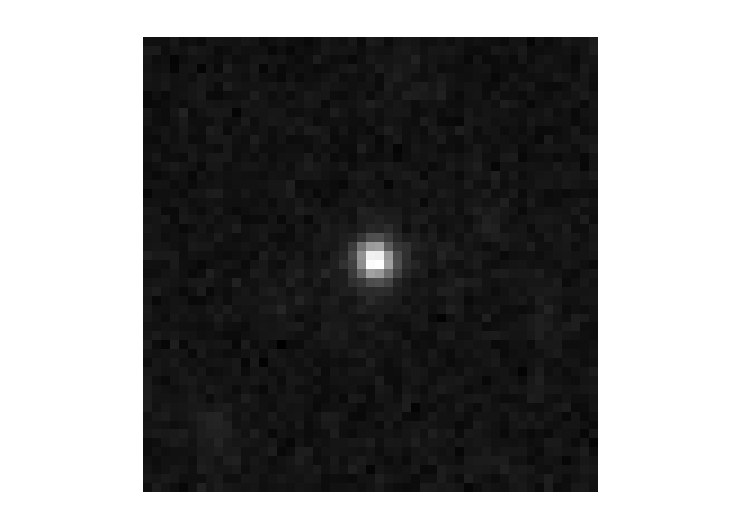}
    \caption{Stacked image of ID 11. The most influential feature is \texttt{Con}.}
    \label{fig:anomaly_11}
\end{figure}

\begin{figure}[H]
    \centering
    \includegraphics[width=1.0\linewidth]{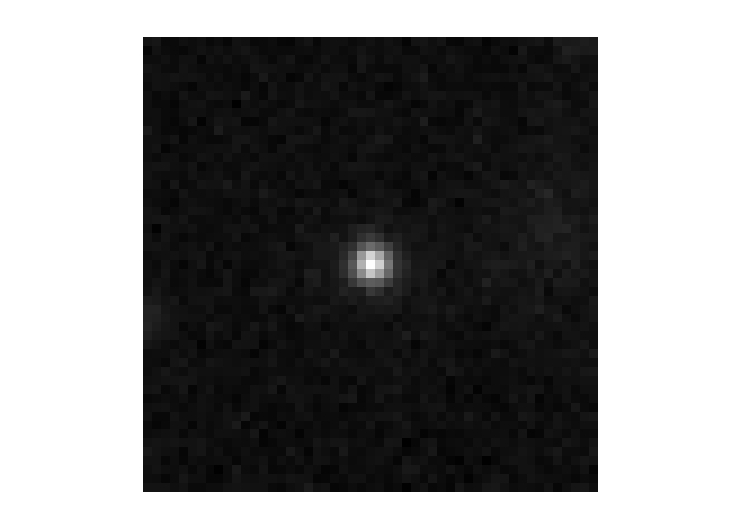}
    \caption{Stacked image of ID 32. The most influential feature is \texttt{GP\_DRW\_$\sigma$}.}
    \label{fig:anomaly_32}
\end{figure}

\begin{figure}[H]
    \centering
    \includegraphics[width=1.0\linewidth]{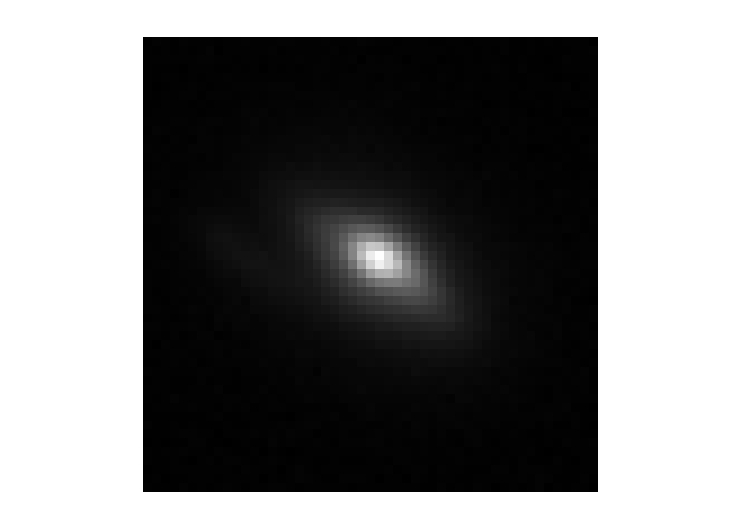}
    \caption{Stacked image of ID 191. The most influential feature is \texttt{Period\_fit\_v2}.}
    \label{fig:anomaly_191}
\end{figure}

\begin{figure}[H]
    \centering
    \includegraphics[width=1.0\linewidth]{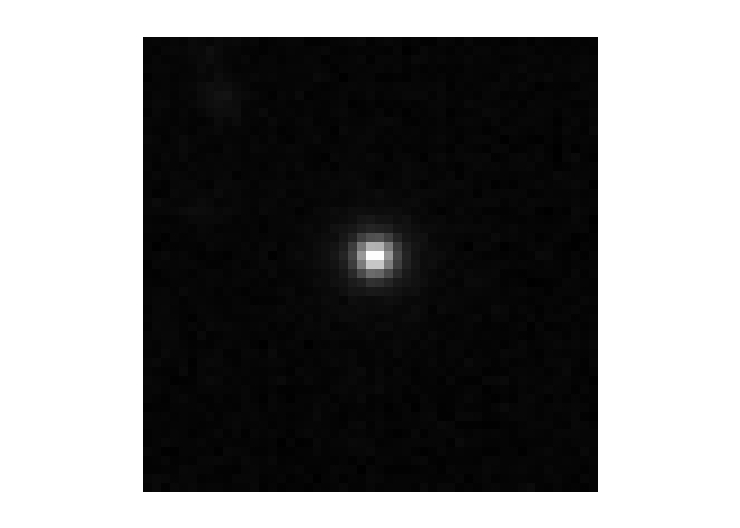}
    \caption{Stacked image of ID 211. The most influential feature is \texttt{GP\_DRW\_$\sigma$}.}
    \label{fig:anomaly_211}
\end{figure}

\begin{figure}[H]
    \centering
    \includegraphics[width=1.0\linewidth]{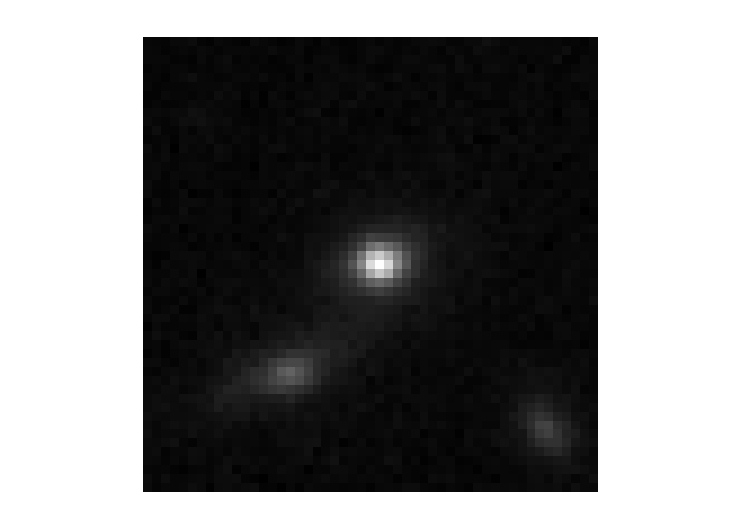}
    \caption{Stacked image of ID 219. The most influential feature is \texttt{Con}.}
    \label{fig:anomaly_219}
\end{figure}

\begin{figure}[H]
    \centering
    \includegraphics[width=1.0\linewidth]{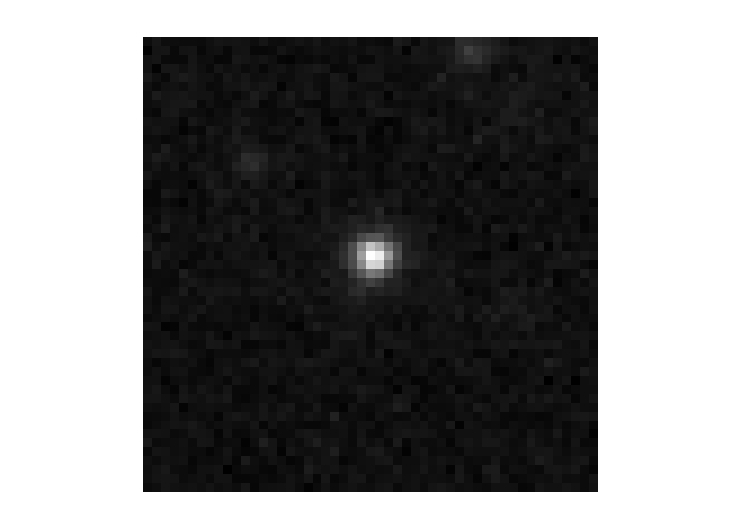}
    \caption{Stacked image of ID 340. The most influential feature is \texttt{PeriodLS\_v2}.}
    \label{fig:anomaly_340}
\end{figure}

\begin{figure}[H]
    \centering
    \includegraphics[width=1.0\linewidth]{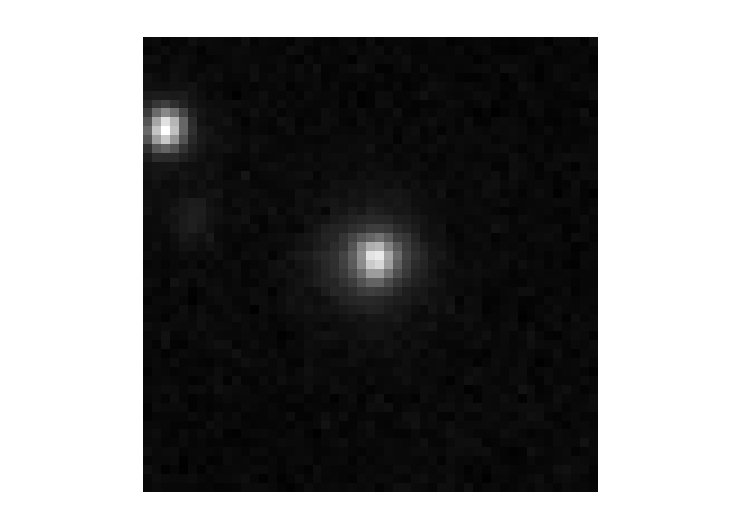}
    \caption{Stacked image of ID 380. The most influential feature is \texttt{Beyond1Std}.}
    \label{fig:anomaly_380}
\end{figure}

\end{appendix}

\end{document}